\documentclass[reprint, aps, prb, superscriptaddress]{revtex4-2}  % for review and submission
\usepackage{chemformula} % Formula subscripts using \ch{}
\usepackage[utf8]{inputenc}
\usepackage[T1]{fontenc}

\usepackage{graphicx}  % needed for figures
\usepackage{dcolumn}   % needed for some tables
\usepackage{bm}        % for maths
\usepackage{amssymb}   % for maths
\usepackage{rotating}
\usepackage{amsmath}   % ditto, for maths
\usepackage{ifthen}
\usepackage{url}
\usepackage{xcolor}
\usepackage[pdfencoding=auto]{hyperref}
\hypersetup{colorlinks=true, citecolor=blue, urlcolor=blue, linkcolor=blue, final}
\usepackage{booktabs}
\usepackage{subcaption}
\usepackage{verbatim}

\usepackage[display-foreign=false]{acro}  
\DeclareAcronym{DFT}{
short = DFT ,
long = density-functional theory
}

\DeclareAcronym{SOAP}{
short = SOAP ,
long = Smooth Overlap of Atomic Positions
}

\DeclareAcronym{AIRSS}{
short = AIRSS,
long = \textit{ab initio} Random Structure Searching
}

\DeclareAcronym{RSS}{
short = RSS,
long = Random Structure Searching
}

\DeclareAcronym{NIB}{
short = NIB,
long = Na-ion battery,
short-plural-form = NIBs,
long-plural-form = Na-ion batteries
}

\DeclareAcronym{NNP}{
short = NNP,
long = Neural Network Potential,
short-plural-form = NNPs
}

\DeclareAcronym{GA}{
short =GA,
long = Genetic Algorithm
}

\DeclareAcronym{K-PCA}{
short =K-PCA,
long = Kernel-Principal Component Analysis
}

\DeclareAcronym{DBSCAN}{
short =DBSCAN,
long = DBSCAN
}

\DeclareAcronym{MLIP}{
short = {MLIP},
long = {machine-learned interatomic potential},
short-plural-form = {MLIPs}
}

\DeclareAcronym{ELF}{
short = {ELF},
long = {Electron Localisation Function},
}

\DeclareAcronym{XAS}{
short = {XAS},
long = {X-Ray Absorption Spectroscopy},
}

\DeclareAcronym{PXRD}{
short = {PXRD},
long = {Powder X-Ray Diffraction},
}

\begin{document}

\title{Structure prediction of stable sodium germanides at 0 and 10 GPa}%https://www.overleaf.com/project/5db84aa215c2600001f6a1ed

\author{James P. Darby}
\email{jpd47@cam.ac.uk}
\affiliation{Engineering Laboratory, University of Cambridge, Cambridge, CB2 1PZ UK}
  
\author{Angela F. Harper} 
\affiliation{Theory of Condensed Matter Group, Cavendish Laboratory,
  J.~J.~Thomson Avenue, Cambridge CB3 0HE, United Kingdom}
\affiliation{Fritz-Haber-Institut der Max-Planck-Gesellschaft, Faradayweg 4-6, 14195 Berlin, DE}  

\author{Joseph R.~Nelson} \affiliation{Department of Materials Science
  and Metallurgy, University of Cambridge, 27 Charles Babbage Road,
  Cambridge CB3 0FS, United Kingdom} \affiliation{Advanced Institute
  for Materials Research, Tohoku University, 2-1-1 Katahira, Aoba,
  Sendai, 980-8577, Japan}

\author{Andrew J.~Morris}
\affiliation{School of Metallurgy and Materials, University of
  Birmingham, Edgbaston, Birmingham B15 2TT, United Kingdom}

\vskip 0.25cm
\date{\today}

\begin{abstract}
In this work we used \textit{ab-initio} random structure searching (AIRSS) to carry out a systematic search for crystalline Na-Ge materials at both 0 and 10~GPa. 
The high-throughput structural relaxations were accelerated using a \ac{MLIP} fit to \ac{DFT} reference data, allowing$\sim$1.5 million structures to be relaxed.
At ambient conditions we predict three new Zintl phases, \ch{Na3Ge2}, \ch{Na2Ge} and \ch{Na9Ge4}, to be stable and a number of Ge-rich layered structures to lie in close proximity to the convex hull. 
The known Na$_\delta$\ch{Ge34} clathrate and \ch{Na4Ge13} host-guest structures are found to be relatively stabilized at higher temperature by vibrational contributions to the free energy.
Overall, the low energy phases exhibit exceptional structural diversity, with the expected mixture of covalent and ionic bonding confirmed using the \ac{ELF}.
The local Ge structural motifs present at each composition were determined using \ac{SOAP} descriptors and the Ge-K edge was simulated for representatives of each motif, providing a direct link to experimental \ac{XAS}.
Two Ge-rich phases are predicted to be stable at 10 GPa;  \ch{NaGe3} and \ch{NaGe2} have simple kagome and simple hexagonal Ge lattices respectively with Na contained in the pores. \ch{NaGe3} is isostructural with the \ch{MgB3} and \ch{MgSi3} family of kagome superconductors and remains dynamically stable at 0 GPa.  Removing the Na from \ch{NaGe2} results in the hexagonal lonsdalite Ge allotrope, which has a direct band gap. 
\end{abstract}


\maketitle

\section{Introduction}
Compounds in the sodium-germanide binary exhibit exceptionally diverse chemistry, with known phases ranging from complex covalently bonded Ge networks to Zintl phases containing various Ge polyanions. This structural diversity gives rise to a wide range of technologically interesting properties. For instance, Zintl phases often have desirable thermoelectric properties \cite{kauzlarich2007zintl}, Na ions can intercalate into various forms of Ge \cite{sharma2018amorphous, chou2015comparative, nandakumar2023type} making it a viable anode for Na-ion batteries, whilst leaching Na from complex binary compounds offers a potential synthesis route to novel Ge allotropes \cite{zhang2016silicon, zhang2021predicted}.
Despite this, and extensive computational investigations into similar systems \cite{zhang2016silicon, zhang2021predicted, morris2014thermodynamically, stratford2017investigating, mayo2017structure}, to date there has been no systematic structure prediction study on the sodium-germanide binary. In this work we use a combination of approaches to search for new phases at 0 and 10 GPa.

\ch{Na4Ge13} was the first known example of a host-guest structure with discrete Zintl polyanions (\ch{Ge4^{4-}}) confined within a three-dimensional covalent framework. This structure was initially reported as Na$_{1-x}$Ge$_{3+z}$ \cite{beekman2010structure} but recent single-crystal diffraction data \cite{stefanoski2018zintl} has confirmed the \ch{Na4Ge13} stoichiometry.

Finally, Na$_\delta$\ch{Ge34}, $\delta=1-3$, is a type-II Ge clathrate with Na intercalated within the Ge cages. The ability to intercalate Na makes this clathrate an interesting candidate anode for \acp{NIB}, which are being actively developed as a promising alternative to Li-ion batteries. 
The low cost, relative abundance and geographically uniform distribution of Na makes \acp{NIB} highly appealing \cite{vaalma2018cost}, particularly for applications such as home storage, where maximising volumetric capacity is less important. However, unlike Li, Na-ions do not intercalate into graphite \cite{stevens2001mechanisms} so finding alternative anodes is crucial. 
There have been extensive investigations into Ge as an alternative, although unfortunately Na-ions cannot be intercalated into crystalline diamond Ge either \cite{sharma2018amorphous}. However they can be intercalated into amorphous bulk \cite{chou2015comparative}, thin films \cite{baggetto2013germanium}, amorphous nano-structures \cite{abel2013nanocolumnar, lu2016germanium} and the aforementioned Na$_\delta$\ch{Ge34} clathrate \cite{nandakumar2023type}. This is due to the more open Ge frameworks in these materials lowering the activation barrier required for Na-ion diffusion. 
Whilst this study will focus exclusively on crystalline NaGe phases, we expect the same low energy local structural motifs to also be present within amorphous materials of similar composition. As such, we include a detailed analysis of these motifs in anticipation that they may be useful when studying amorphous Ge anodes. 

The searches at 10 GPa were inspired by a sequence of studies carried out on the similar LiSi \cite{zhang2016silicon}, NaSi \cite{zhang2021predicted} and MgSi \cite{zha2023refined} systems. The primary aim of these works was to discover stable Si-rich structures comprised of metal atoms sitting within a covalently bonded silicon network. The metal atoms could then be removed via thermal degassing \cite{kim2015synthesis} at lower pressure, hopefully leaving behind a stable Si framework and hence discovering a new Si allotrope. 
The discovery of new Group XIV allotropes is particularly interesting not only because of their potential use as novel anodes, but also for their possible optical properties.
Integrating optically active devices into microelectronic chips is highly desirable but challenging, because the new device must be both optically active and chemically compatible with the Si chip. At present, III-IV semiconductors are used but this requires complex heterogengous integration \cite{miller2009device}. In contrast, cubic diamond Si and Ge are exceptionally chemically compatible with the silicon substrate but have indirect band gaps so cannot emit light efficiently. As such, novel Si and Ge allotropes with direct band gaps are highly sought after. Hexagaonal Ge in the lonsdalite structure has a direct band gap but the associated optical transition is exceptionally weak, leading to various efforts at improving the transition strength through alloying \cite{borlido2023ensemble}, applying strain \cite{suckert2021efficient} and doping \cite{cai2021carrier}. 

In this work a \ac{MLIP} was trained and used to accelerate crystal structure prediction of binary sodium-germanide phases at 0 and 10 GPa. Computed phonon frequencies were used to assess both the dynamical stability and thermodynamic stability of low energy structures at elevated temperatures. See Section \ref{sec:methods} for more details. 
At 0 GPa we predict three new phases \ch{Na3Ge2}, \ch{Na2Ge} and \ch{Na9Ge4} to be stable or within 3 meV/atom of the convex hull and a handful of new metastable Ge-rich layered structures. At 10 GPa all predicted structures  are new, including \ch{NaGe2} and \ch{NaGe3} which consist of Na atoms contained within simple hexagonal and simple kagome Ge lattices respectively. The results are presented in Section \ref{sec:results} with conclusions following in Section \ref{sec:conclusion}.

\section{Methods} \label{sec:methods}

\subsection{Structure Searching}

Firstly, the four known Na-Ge binary structures, \ch{NaGe}, \ch{NaGe34}, \ch{Na4Ge13} and \ch{Na12Ge17}, were extracted from the ICSD~\cite{ICSD}. The latter three are all reported with some level of structural disorder, ranging from fractional occupancy across Na sites to orientational disorder of \ch{Ge4} tetrahedra, so various ordered variants were investigated. Secondly, all binary structures in the ICSD comprised of group~I and group~XIV elements were extracted and used as prototypes to construct the corresponding Na-Ge structure via element-swapping group~I$\rightarrow$Na and group~XIV$\rightarrow$Ge. These structures were then relaxed to local enthalpy minima, at both 0 and 10 GPa, using \ac{DFT}

Relaxing known prototypes is a fast and simple way of obtaining an initial set of chemically reasonable structures. Moreover, this set is helpful in checking that more advanced structure prediction algorithms are providing an adequate coverage of the potential energy surface as, at a minimum, they should recover the low-energy prototypes. One such algorithim is \ac{AIRSS} \cite{schusteritsch2014predicting}, which 
involves generating random initial structures, subject to physically reasonable constraints, and then relaxing them to local enthalpy minima using \ac{DFT}. Whilst exceptionally simple, \ac{AIRSS} has proven successful across a wide range of structure prediction tasks including high-pressure hydrides \cite{pickard2006high, pickard2007metallization}, grain-boundary interfaces \cite{schusteritsch2014predicting}, battery anodes \cite{mayo2016ab, harper2020computational}, 2D phases of ice \cite{chen2016two}, 1D nanowires \cite{wynn2017phase} and even metal-organic frameworks \cite{xu2023experimentally}.   Furthermore, the relaxation of independent trial structures is trivially parallelisable, allowing for efficient utilisation of HPC resources. 

At both pressures, \ac{AIRSS} was used to generate $\sim$60,000 structures in the composition range Na$_{15}$Ge-NaGe$_{15}$. Structures were generated with a maximum of 24 atoms per primitive unit cell and using, typically 2-6, randomly chosen symmetry operations. The minimum distance between atoms and unit cell volumes were also constrained to bias the search towards physically reasonable structures. 
After relaxation, the 0 GPa AIRSS results contained many low energy structures with Ge-Ge dumbbells and Ge$_4$ tetrahedra. As such, an additional $\sim$3000 structures were generated using these motifs as the base structural units, similar to the approach taken in ref. \cite{ahnert2017revealing}. For these searches, symmetry constraints were applied using the Wyckoff-Aligned Molecules approach \cite{darby2020ab} which allowed the units to occupy high symmetry Wyckoff positions.

Analysis of the relaxed structures indicated that good coverage of the potential energy surface was achieved at 10 GPa but, interestingly, the same was not true at 0~GPa. This was particularly evident in the Ge-rich region, see Figure~\ref{sfig:0GPa_search_hull}, where, despite the existence of known phases, very few energetically competitive structures were found.
Given this, \ac{AIRSS} was used to generate an additional one million structures with larger unit cells, up to 40 atoms per primitive unit cell, in the, slightly extended, composition range \ch{NaGe19}-\ch{Na19Ge}. 
This increased scale of searching was made feasible by relaxing the structures using a \ac{MLIP}, see Section \ref{section:NNP} for details, which reduced the cost of a typical relaxation by many orders of magnitude. Using \acp{MLIP} to accelerate structure prediction in this way is now common practice \cite{podryabinkin2019accelerating, bernstein2019novo, pickard2022ephemeral} and many different forms of \ac{MLIP} can be used.

Further searches were also performed using the \ac{MLIP} coupled with the \texttt{illustrado} \cite{illustrado} \ac{GA}. 
These were done in the restricted composition range Na$_{19}$Ge-NaGe to specifically target Ge-rich structures containing complex Ge networks. 
The \ac{GA} was seeded using low energy structures from the previous searches and used to generate a total of $\sim$100,000 further structures. 
A ``continuous-birth'' strategy, where child structures are generated continuously rather than in distinct batches demarcating generations, was used to make efficient use of computational resources. 
\texttt{illustrado} uses simple cut-and-splice cross-over as well as various mutation operations to generate child structures from the existing population. 

Finally, all structures relaxed with the \ac{MLIP} that were within 20 meV/atom of the convex hull were re-optimised with \ac{DFT}, using the same settings as for the initial \ac{AIRSS}.

\subsection{Neural Network Potential} \label{section:NNP}

The \texttt{aenet} software \cite{artrith2016implementation, artrith2017efficient} was used to train a \ac{MLIP} using $\sim$119,000 randomly chosen snapshots from the structural relaxations performed in the initial \ac{AIRSS} at 0 GPa. This \ac{MLIP} was then used to perform a trial \ac{AIRSS} search consisting of $\sim$250,000 structures. While this search was promising - a number of known structures were found  - it revealed some shortcomings with the \ac{MLIP}. Firstly, five very dense structures were predicted to be exceptionally stable; these structures were $\sim3\times$ denser and $\sim$1 eV/atom more stable than comparable known phases. Secondly, a number of Ge-rich, Ge$_x$Na with $x$ > 7,  structures were predicted to be overly stable by $\sim$250 meV/atom. 

We attributed both of the cases described above to extrapolation outside of the training dataset. 
Whilst these cases were easy to spot manually we desired a more automated approach for identifying extrapolation.
To this end, we trained three additional \acp{MLIP}, using different randomly chosen snapshots, and combined them with the original potential to form a committee \cite{van2023hyperactive, peterson2017addressing, smith2018less}.
Evaluating the committee on random snapshots taken from the 10 GPa  \ac{AIRSS}, chosen so that no additional \ac{DFT} calculations were required, revealed a useful correlation between the variance of the committee energy prediction and the energy error, see Figure \ref{fig:ensemble_10GPa}. 
Given this, the committee variance was used to select $\sim$8,500 structures from the trial \ac{MLIP}-accelerated \ac{AIRSS} which were then added to the training set. 
The mean absolute energy error of the committee across these structures was $\sim$1 eV/atom, compared to 16 meV/atom on the test set, indicating that the selection criterion significantly outperformed adding further random snapshots.
Inspection of the new structures revealed a strong bias towards Ge-rich structures, see Figure \ref{fig:plus_stoich}, with complex networks and those with extreme densities.
Additionally, Na-Na, Na-Ge and Ge-Ge dimers with separations of 0.6-5.8~\AA~were used to augment the training set. 
This was done in an attempt to make the structural relaxation of unusual configurations, which occur frequently during \ac{AIRSS}, more robust.

A new \ac{MLIP} was then trained on the final dataset containing $\sim$128,000 configurations. A comparison between the energies and forces predicted by the \ac{MLIP} and the reference method for an independent test set are show in Figure \ref{fig:NNP_performance}. The test set was formed by taking a single snapshot from 3500 distinct structural relaxations, all of which were fully excluded from the training set. Only a single snapshot was taken from each trajectory to ensure the test configurations were not correlated. The mean absolute energy and force  errors on the test set were 16 meV/atom and 0.11 eV/\AA~ respectively. The final dataset and potential are available on Zenodo \cite{zenodo} whilst details of the \ac{MLIP} architecture are given in the Supplemental Material.

\subsection{DFT}

All \ac{DFT} calculations were performed using the CASTEP \cite{clark2005first} plane-wave pseudo-potential code and the PBE exchange-correlation functional \cite{perdew1996generalized}.
During AIRSS, the structural relaxations were performed using a plane-wave cutoff of 400 eV, k-point spacing of 2$\pi\times0.05$~\AA$^{-1}$ and Vanderbilt ultra-soft pseudo-potentials \cite{vanderbilt1990soft}. 
All structures within 15 meV/atom (20 meV/atom for 10 GPa) of the convex hull were then re-optimised at a higher level of accuracy - plane-wave cutoff of 800 eV, k-point spacing of 2$\pi\times0.03$~\AA$^{-1}$ and the  ``C19'' CASTEP on-the-fly ultra-soft pseudo-potentials.
With these parameters energies, forces and stresses were converged to within 1 meV/atom, 0.005 eV/\AA~and 0.03 GPa respectively.

Phonon calculations were then performed to asses the dynamical stability of low-energy structures using the finite-displacement approach with non-diagonal supercells \cite{lloyd2015lattice} and a phonon $q$-point spacing of 2$\pi\times0.1$~\AA$^{-1}$, corresponding to a single supercell with dimension at least 10 x 10 x 10~\AA. For some structures a finer grid of 2$\pi\times0.0625$~\AA$^{-1}$ was required to converge the bandstructure.
The \texttt{SeeK-path} \cite{hinuma2017band, togo2018texttt} package was used to select the path for the phonon band-structure calculations whilst a finer $q$-point spacing of 2$\pi\times0.01$~\AA$^{-1}$ was used for the Fourier interpolation when computing both the phonon band-structure and density of states.
Imaginary frequencies were observed in the phonon bandstructures of \ch{NaGe2} and \ch{Na9Ge4} (0 GPa) and \ch{Na3Ge4} (10 GPa) so these structures were perturbed along the corresponding modes and re-relaxed. 
The phonon frequencies were then used to compute the vibrational contribution to the free energy \cite{baroni2001phonons} within the harmonic approximation. The computed free energies were then used to construct the temperature-dependent convex-hull.

Electronic bandstructures were computed with and without spin-polarisation using a k-point path spacing of 2$\pi\times0.03$~\AA$^{-1}$, whilst 
the \texttt{OptaDOS} code \cite{morris2014optados} was used to project the electronic density of states onto the local atomic orbitals. 

\subsection{Workflow and Analysis}
The \texttt{run3} utility from the open source \texttt{matador} \cite{evans2020matador} python package was used to perform all high-throughput \ac{DFT} calculations. Furthermore, \texttt{matador} was also used to fetch prototype structures from the ICSD, identify duplicate structures, compute the convex hulls and for all plotting. The convex hulls were computed using
\begin{equation}
    E_\mathrm{f}(\mathrm{Na}_x\mathrm{Ge}_y) = E(\mathrm{Na}_x\mathrm{Ge}_y) - x E(\mathrm{Na}) - y E(\mathrm{Ge}),
\end{equation}
where $E(\mathrm{Na})$ and $E(\mathrm{Ge})$ are the appropriate chemical potentials for Na and Ge respectively. We considered the Im$\bar{3}$m, R$\bar{3}$m and Fm$\bar{3}$m allotropes of Na, finding the Im$\bar{3}$m (bcc) form to be lowest in energy at both pressures. For Ge, the $\alpha, \beta, \gamma, \delta, \epsilon$ and 4H allotropes, as summarised in  ref. \cite{fassler2007germanium}, were assessed with $\alpha$-Ge (Fd$\bar{3}$m,  diamond structure) found to be stable at 0 GP and $\beta$-Ge (I$4_1$/amd, $\beta$-Sn structure) stable at 10 GPa, consistent with experimental observation.

The local motif analysis was performed using the \ac{SOAP} power spectrum \cite{bartok2013representing} as an atomic descriptor. Only Ge atoms were considered when computing the atomic descriptors. \ac{K-PCA}  \cite{scholkopf1997kernel} was used to perform dimensional reduction whilst \ac{DBSCAN} \cite{ester1996density}  was used to automatically identify clusters; both algorithms were used through the \texttt{scikit-learn} python package \cite{pedregosa2011scikit}.
All data analysis was performed using the analysis notebook on Zenodo \cite{zenodo}.

\section{Results} \label{sec:results}
Structures are reported as new if not present in either the OQMD \cite{saal2013materials}, ICSD \cite{ICSD} or Materials Project \cite{jain2013commentary} and if, to the best of our knowledge, they have not been experimentally synthesised. We also note that of the 384,872 structures predicted to be stable in the recent study using GNoME \cite{merchant2023scaling}, 184 contain both Na and Ge of which 12, 115 and 57 contain 3, 4 and 5 elements respectively. That no novel Na-Ge binary phases were reported suggests that the benefits of careful, directed searching should not be overlooked.

\subsection{0 GPa}

\begin{figure*}[htb!]
  \includegraphics[width=\textwidth]{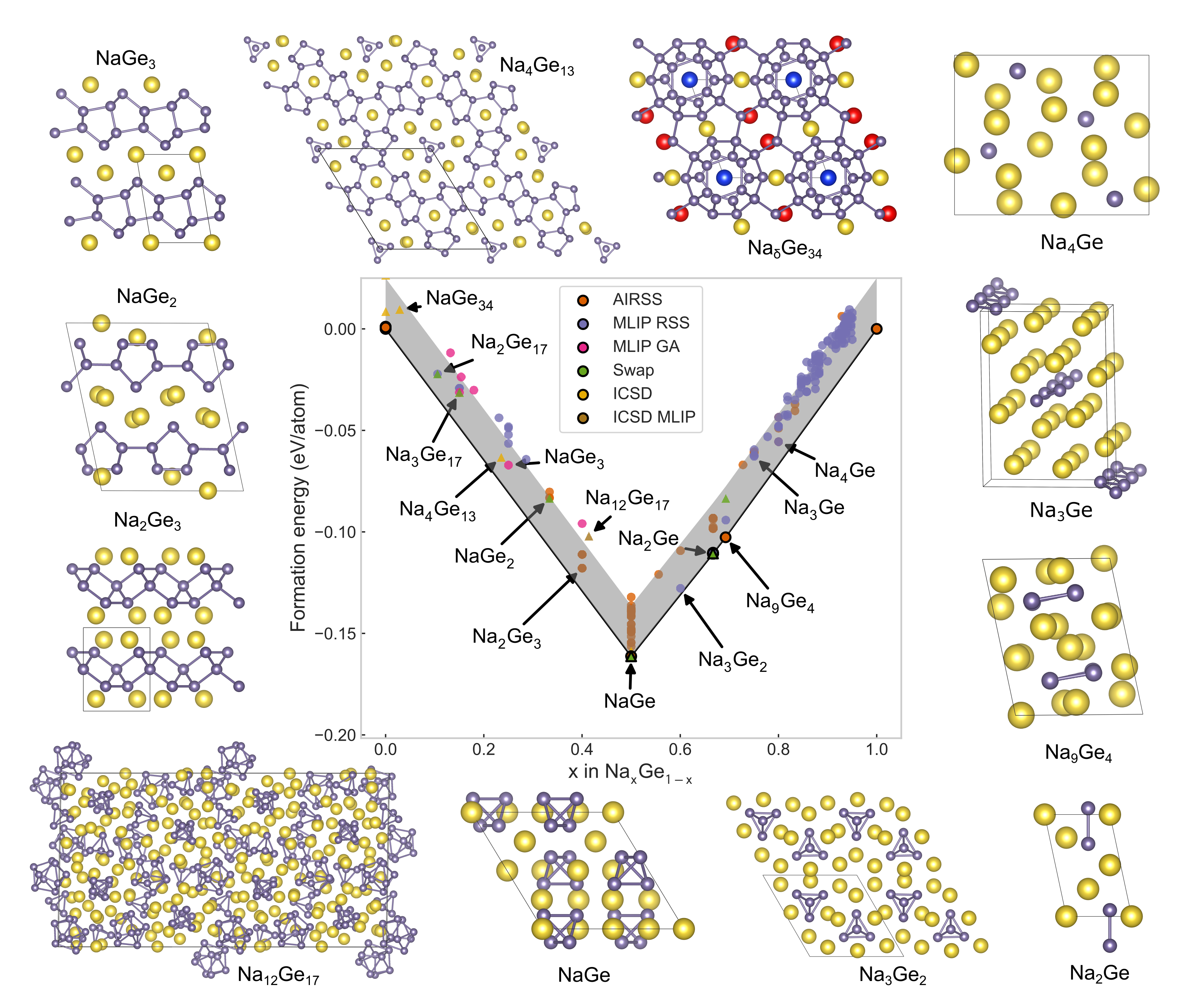}
  \caption{The 0 GPa Na-Ge convex hull, computed using the accurate DFT settings with bcc-Na and diamond-Ge as the chemical potentials. The grey band indicates +20 meV/atom whilst Ge-Ge bonds are drawn between atoms separated by 2.8 \AA~or less. \ch{NaGe34}, \ch{Na2Ge17} and \ch{Na3Ge17} all share the same Ge clathrate structure so are shown as Na$_\delta$\ch{Ge34} with the Na colored blue, blue and yellow and blue, yellow and red for $\delta$=1, 4 and 6 respectively. Markers are colored according to the provenance of the corresponding structure where ``MLIP RSS'' and ``MLIP GA'' indicate structures found using the \ch{MLIP} and then optimised with \ac{DFT}. Due to it's large size, the hull distance for \ch{Na12Ge17} was computed using the \ac{MLIP} to be 32 meV/atom and it is shown in this postion. }
  \label{fig:0GPa_Hull}
\end{figure*}

\begin{table}[h!]
	\centering
	\caption[Summary of low energy NaGe binary phases at 0 GPa]{Na-Ge phases within 25 meV/atom of the 0~GPa convex hull are listed. Compositions calculated to lie on the convex hull are shown in \textbf{bold}. Na-rich structures beyond Na$_4$Ge are not included for brevity. \ch{NaGe34}, \ch{Na2Ge17} and \ch{Na3Ge17} were derived from the disordered ICSD structure 192631. Space groups computed using spglib with symmetry tolerance of 0.01. $^*$\ch{Na12Ge17} is an ordered variant of ICSD structure 412932. It contains 464 atoms so the hull distance was estimated using the NNP rather than DFT.  $^\dag$ \ch{NaGe2} is derived from a swap from ICSD structure 170467 then perturbing along an imaginary phonon frequency.}
	\begin{tabular}{ccccc}
            \toprule
		formula & $\Delta$ E (meV/atom) & Z & space group & provenance	\\
            \midrule
		\textbf{$\alpha$-Ge} & - &  2 & Fd$\bar{3}$m & ICSD 43422 \cite{cooper1962precise}  \\
            4H-Ge & 9 &  8 & P6$_3$/mmc & ICSD 167204 \cite{kiefer2010synthesis}  \\
            $\epsilon$-Ge & 26 &  34 & Fd$\bar{3}$m & ICSD 245951 \cite{schwarz20083d}  \\
		\ch{NaGe34} & 19 &  1 & R$\bar{3}$m & Swap 192631 \cite{bohme2014preparation} \\
		\ch{Na2Ge17} & 12 &  2 & Fd$\bar{3}$m &  Swap 192631 \cite{bohme2014preparation} \\
		\ch{Na3Ge17} & 17 &  2 & P$\bar{1}$ &  Swap 192631 \cite{bohme2014preparation}\\
		\ch{Na4Ge13} & 13 &  4 & P3 & ICSD 256042 \cite{stefanoski2018zintl} \\
		\ch{NaGe3} & 14 &  2 & C2/m & NNP GA  \\
		\ch{NaGe2} & 24 &  16 & P2$_1$/m & $^\dag$ICSD 170467 \cite{dubois2005nasn2} \\
		\ch{Na2Ge3} & 11 &  4 & P$\bar{4}$2$_1$m & AIRSS \\
            \ch{Na12Ge17} & $^*$32 &  16 & P2$_1$/c & ICSD 412932 \cite{carrillo2003na12ge17} \\
		\textbf{\ch{NaGe}} & - &  16 & I4$_1$/acd & Swap 641294  \cite{hewaidy1964struktur}\\
             & 3 &  16 & P2$_1$/c & ICSD 193521  \cite{morito2015crystal}\\
		\ch{Na5Ge4} & 24 &  2 & Pbam & NNP RSS \\
		\ch{Na3Ge2} & 3 & 4 & P31c & NNP RSS \\
		\textbf{\ch{Na2Ge}} & - & 2 & R$\bar{3}m$ & Swap 24146 \cite{axel1965kristallstruktur} \\
		\textbf{\ch{Na9Ge4} }& - & 2 & P$\bar{1}$ & NNP RSS \\
		\ch{Na8Ge3} & 24 & 1 & P1 & AIRSS \\
		\ch{Na3Ge} & 20 & 4 & Pnma & NNP RSS \\
		\ch{Na7Ge2} & 21 & 1 & Cmmm & NNP RSS \\
		\ch{Na4Ge} & 11 & 4 & Pnma & AIRSS \\
            %pasted in with correct symmetry tolerance, maybe show in SI?
            % \ch{Ge3Na13} & 0.0199 &  1 & Cm &  \\
            % \ch{Ge2Na9} & 0.0173 &  1 & P2/m &  \\
            % \ch{GeNa5} & 0.0151 &  2 & P2_1/m &  \\
            % \ch{Ge3Na16} & 0.0208 &  1 & P321 &  \\
            % \ch{Ge2Na11} & 0.0225 &  1 & P-1 &  \\
            % \ch{GeNa6} & 0.0165 &  2 & P2_1/m &  \\
            % \ch{Ge2Na13} & 0.0206 &  1 & P2/m &  \\
            % \ch{GeNa7} & 0.0120 &  2 & P2_1/m &  \\
            % \ch{Ge2Na15} & 0.0170 &  1 & P-1 &  \\
            % \ch{GeNa8} & 0.0233 &  2 & P2_1/m &  \\
            % \ch{GeNa9} & 0.0173 &  2 & P2_1/m &  \\
            % \ch{GeNa10} & 0.0225 &  2 & P2_1/m &  \\
            % \ch{GeNa11} & 0.0181 &  2 & P-1 &  \\
            % \ch{GeNa12} & 0.0207 &  2 & R-3m &  \\
            % \ch{GeNa13} & 0.0166 &  2 & P-1 &  \\
            % \ch{GeNa14} & 0.0150 &  2 & P-1 &  \\
            % \ch{GeNa15} & 0.0157 &  2 & P-1 &  \\
            % \ch{GeNa16} & 0.0208 &  1 & P-1 &  \\
            % \ch{GeNa17} & 0.0236 &  1 & P-1 &  \\
            % \ch{GeNa18} & 0.0155 &  2 & Pmmn &  \\
            % \ch{GeNa19} & 0.0114 &  2 & I4/m &  \\
            % \ch{Na} & 0.0000 &  1 & Im-3m &  \\
		\textbf{Na} & - & 1 & Im$\bar{3}$m & ICSD 44757 \cite{barrett1956x} \\
  \bottomrule
	\end{tabular}	
	\label{table:NaGe_0GPa}
\end{table}

The only previously known binary phase unambiguously predicted to be stable is NaGe, which is comprised of \ch{Ge4^{4-}} tetrahedral Zintl ions surrounded by positively charged Na ions. As seen in Figure \ref{fig:0GPa_Hull}, there are many different possible packings of these ions that are energetically competitive. We find an $I4_1/acd$ arrangement to be $\sim$ 3 meV/atom lower in energy than the $P2_1/c$ structure reported previously \cite{hewaidy1964struktur}, but note that this difference is well within the expected \ac{DFT} error, and is predicted to decrease at elevated temperatures, see Figure \ref{sfig:Thull_dists}. NaGe is reported as a black powder \cite{ma2009versatile} which is consistent with the semi-conducting character of the computed electronic bandstructure shown in Figure \ref{sfig:NaGe_Na3Ge2_bandstructure}.

The stability and Na-ion diffusion in two of the other known phases \ch{Na4Ge13} and Na$_\delta$Ge$_{34}$ was recently investigated in ref. \cite{nandakumar2023type}. This was done in the context of using the Na$_\delta$Ge$_{34}$ clathrate phases, derived from $\epsilon$-Ge, as an intercalation anode in Na-ion batteries. The cage-like structure of $\epsilon$-Ge is formed from face sharing \ch{Ge20} dodechahedra and \ch{Ge28} hexakaidecahedra, with each cage providing a potential intercalation site for Na. In ref. \cite{nandakumar2023type} they found that the centers of the \ch{Ge20} cages are preferentially occupied before the \ch{Ge28} cages, where the Na sits off-center. We observe the identical behaviour in \ch{NaGe34}, \ch{Na2Ge17} and \ch{Na3Ge17}, see Figure \ref{sfig:Na6Ge34_cages}, but do not predict \ch{Na2Ge17} to lie on the convex hull at 0 K. Instead, we find it to be stabilised by vibrational contributions to the free energy above 500 K, as shown in Figure \ref{sfig:Thull_dists}. We attribute the predicted discrepancy to the different choices of planewave cutoffs.

The same work also predicts \ch{Na24Ge74}, derived from the structural model of Na$_{1-x}$Ge$_{3+z}$ given in ref. \cite{beekman2010structure}, to be stable. We instead consider the more recent \ch{Na4Ge13} model, see Figure \ref{fig:0GPa_Hull}, derived from single-crystal diffraction in ref. \cite{stefanoski2018zintl}. This structure is a rare example of a guest-host structure where \ch{Ge4^{4-}} Zintl irons are confined inside the channels of a 3D covalently bonded Ge network. Specifically we use the ``c$\times2$'' model from ref. \cite{stefanoski2018zintl}, which is the smallest realisation of the structure without rotational disorder of the \ch{Ge4^{4-}} ions. As with the clathrate phases, \ch{Na4Ge13} is increasingly stable at higher temperatures and is predicted to be only $\sim$5 meV/atom above the convex hull at 500 K. 

The final known phase is \ch{Na12Ge17} which contains a 1:2 ratio of \ch{Ge9^{4-}} and \ch{Ge4^{4-}} Zintl ions. Due to the exceptionally large unit cell, 464 atoms, we assessed the stability of this structure using the \ac{MLIP} only, finding it to be $\sim$ 30 meV/atom above the convex hull, which is within the expected accuracy of the \ac{MLIP}. 

\begin{figure}[h!]
  \includegraphics[width=0.25\textwidth]{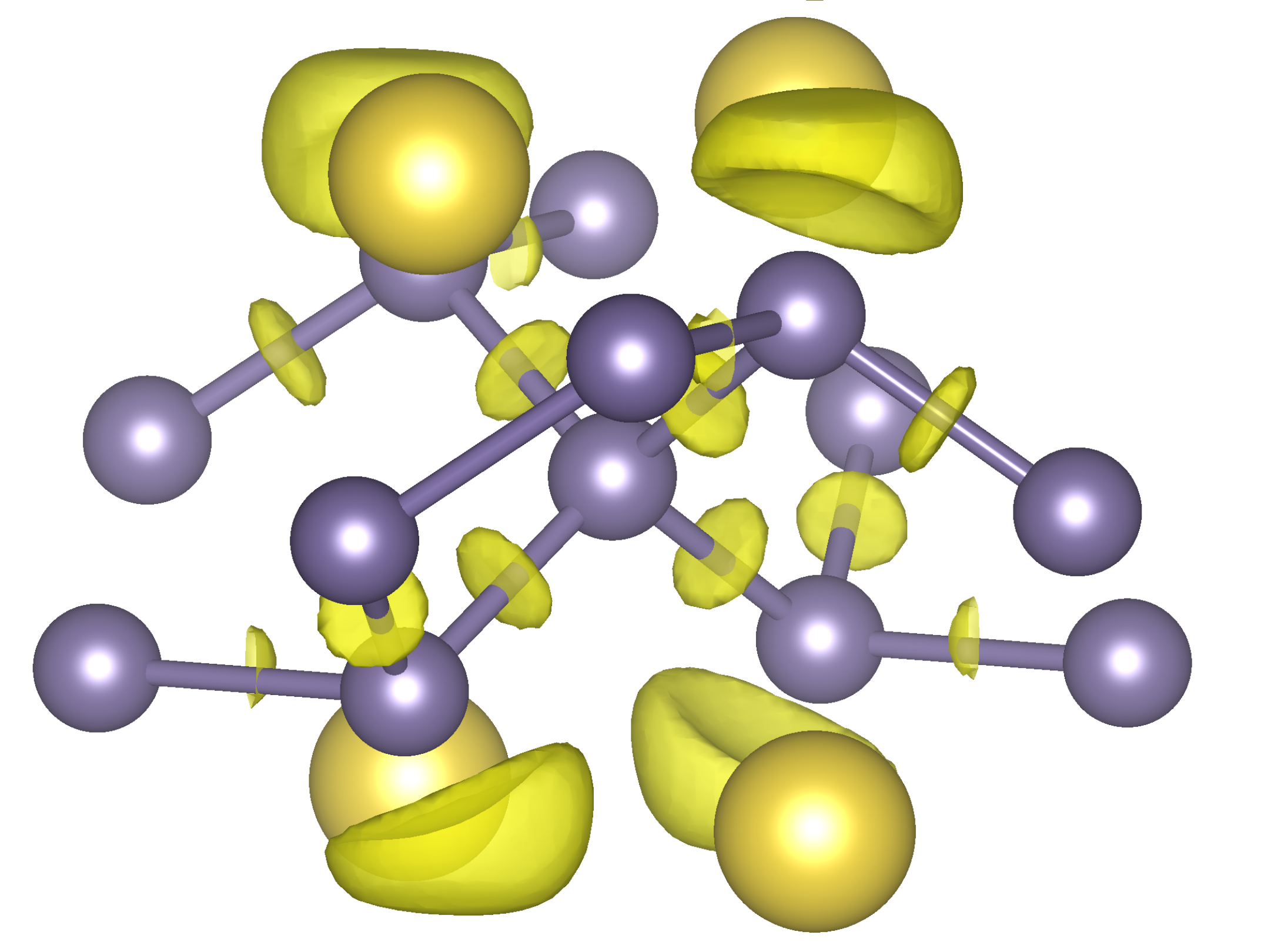}
  \caption{The ELF of \ch{Na2Ge3} is shown with an isosurface value of 0.67. The covalent Ge-Ge bonds are clearly visible whilst a lone pair can be seen on each three-fold coordinated Ge. }
  \label{fig:layered_zintl_0.67}
\end{figure}

\begin{figure}[h!]
  \includegraphics[width=0.48\textwidth]{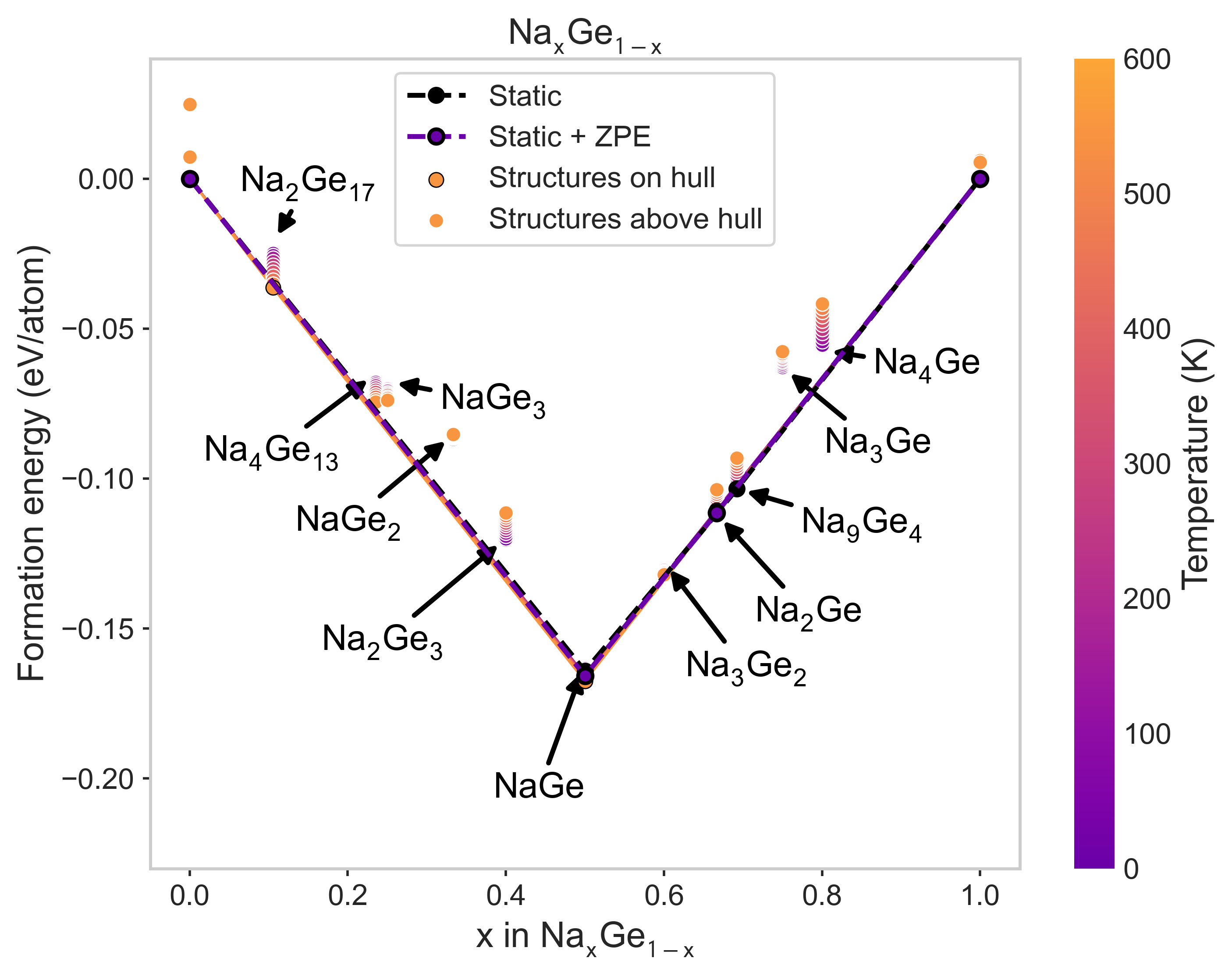}
  \caption{The predicted temperature dependent (harmonic approximation) Na-Ge convex hull at 0 GPa. \ch{Na2Ge17} and \ch{Na4Ge13} are relatively stabilised at higher temperatures.}.
  \label{fig:Thull_0GPa}
\end{figure}

On the Ge-rich side we find three new phases \ch{NaGe3}, \ch{NaGe2} and \ch{Na2Ge3} to be similarly far above the convex hull as the known phases, indicating that they may be synthesisable. All three phases are layered, with 2D Ge networks separated by layers of Na atoms. The Ge networks consist of both 3 and 4-fold coordinate Ge atoms and are all approximately 2-3 Ge atoms in thickness. \ch{NaGe2} is isostructural with \ch{NaSn2} \cite{dubois2005nasn2}, and very similar to \ch{Li7Ge12}, both of which are are Zintl phases with negatively charged 2D Ge layers \cite{dubois2005nasn2, beekman2019zintl}. Inspecting the computed \ac{ELF} for \ch{NaGe3}, \ch{NaGe2} and \ch{Na2Ge3}, shown in Figures \ref{fig:layered_zintl_0.67} and \ref{sfig:ELF_layered_zintl}, supports the Ge-Ge bonding shown in Figure \ref{fig:0GPa_Hull} and indicates that all the 3-fold coordinated Ge atoms have a lone-pair. This is in agreement with the behaviour of the 3-fold coordinated Sn in \ch{NaSn2} and suggests that all three phases can be considered as Zintl-like. Furthermore, the ratio of 3-fold:4-fold coordinated Ge in \ch{NaGe3}, \ch{NaGe2} and \ch{Na2Ge3} is 1:2, 1:1 and 2:1 respectively. This can be understood as each Na atom donating an electron to one of the Ge atoms, which then behaves as a Group XV element like P and so forms three bonds, whilst the remaining Ge atoms form four bonds. As shown in Figure \ref{fig:Thull_0GPa}, \ch{NaGe3} is also the only one of the three phases to remain energetically competitive at elevated temperatures. 

Analysis of the Na-rich structures revealed three new phases all on, or very close to, the convex hull. Of these, \ch{Na2Ge} and \ch{Na9Ge4} both contain negatively charged Ge-Ge dumbbells (Hirshfeld charge of -0.23$e$) surrounded by Na atoms. The dumbbell structural motif is also observed in the analogous lithium-silicides and lithium-germanides \cite{morris2014thermodynamically} with \ch{Na2Ge} having the same structure as \ch{Li2Se} \cite{axel1965kristallstruktur}, whilst \ch{Na9Ge4} adopts a different arrangement to the Cmcm symmetry observed in \ch{Li9Ge4} \cite{hopf1970kristallstruktur} and \ch{Na9Sn4} \cite{midler1978strukturen}. Both phases are predicted to be metallic and remain stable after the inclusion of zero-point energy up to approximately 100 K. 

\begin{figure}[h!]
  \includegraphics[width=0.48\textwidth]{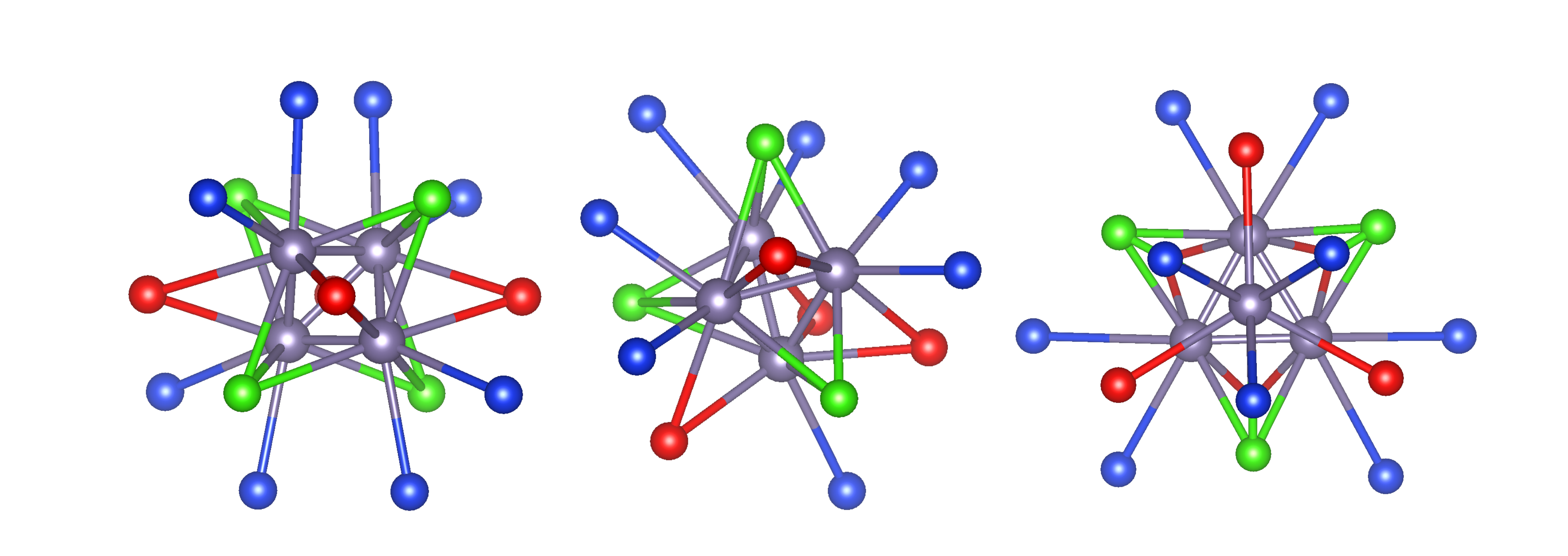}
  \caption{From left to right the coordination of the \ch{Ge4} tetrahedra in I4$_1$/acd NaGe, P2$_1$/c \ch{NaGe} and \ch{Na3Ge2} are shown, where the surrounding Na atoms have been colored green, red and blue if they are adjacent to a face, edge or corner of the tetrahedron respectively. Bonds have been drawn between atoms closer than 3.7 \AA~and are intended to guide the eye, rather than indicate a covalent bond.}
  \label{fig:tet_coordination}
\end{figure}

The predicted \ch{Na3Ge2} structure is also metallic and contains negatively charged tetrahedral \ch{Ge4} ions, as see in both I4$_1$/acd and P2$_1$/c \ch{NaGe}. The strong structural similarity between \ch{Na3Ge2} and \ch{NaGe} hints at the existence of intermediate phases with properties that could be tuned by varying the Na content. As \ch{Na3Ge2} is metallic and \ch{NaGe} has a bandgap of $\sim0.6$ eV, see Figure \ref{sfig:NaGe_Na3Ge2_bandstructure} for comparison, the bandgap would likely be tunable in this way. The large number of energetically competitive packings of \ch{NaGe} found in the \ac{AIRSS} imply that such intermediate phases could be amorphous, particularly given that they would be further stabilised by configurational entropy. Examining the different coordination of \ch{Ge4} tetrahedra shows that Na preferentially sits adjacent to either the face, edge or corners of the \ch{Ge4} tetrahedra, as depicted in Figure \ref{fig:tet_coordination}. Interestingly the coordination is different in all three phases and is notably less symmetric in P2$_1$/c \ch{NaGe}.

We note that previous studies have also reported \ch{Na3Ge} \cite{wang2012thermodynamic, drits1982state} as stable, although we could not find an associated structural model. Both the OQMD \cite{saal2013materials} and Materials Project \cite{jain2013commentary}  provide putative structures for \ch{Na3Ge} with I4/mmm and Fm$\bar{3}m$ symmetry respectively. These structures contain isolated Ge atoms and lie $\sim$70 meV atom above the convex hull whilst the lowest energy structure found in this search contains 1D ``zig-zag'' Ge chains, but is still $\sim$20 meV/atom above the hull. We do however see isolated Ge atoms predicted in \ch{Na4Ge}.

\begin{figure}[h!]
  \includegraphics[width=0.5\textwidth]{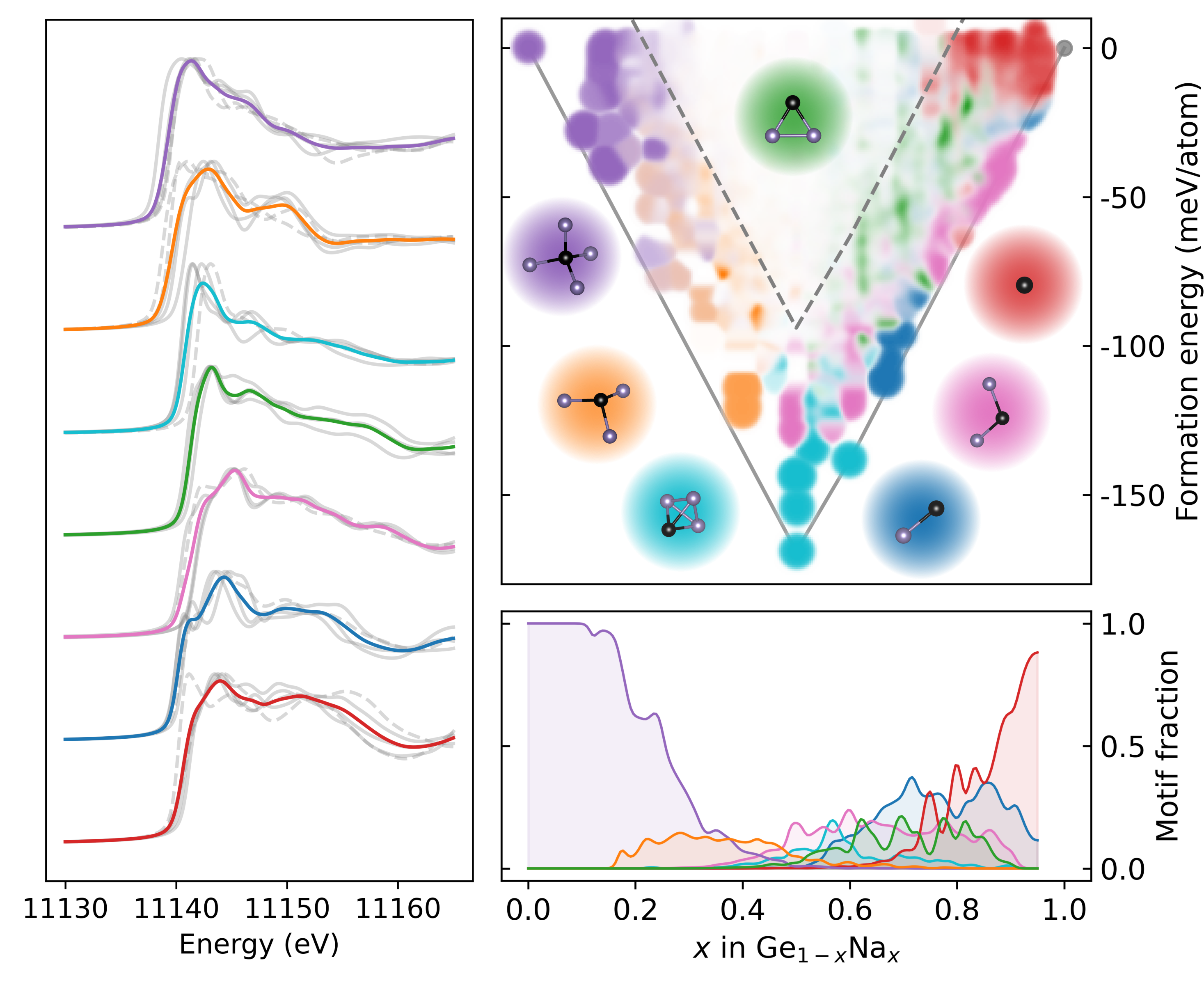}
  \caption{The upper right panel shows the 0 GPa convex hull colored by the fraction of Ge atoms assigned to the local environments shown surrounding the hull, central atom shown in black. The panel below shows the fraction of each motif at each composition. The left panel shows the Ge K-edge XAS spectra computed for 3-5 Ge environments taken from representative structures. The colored line shows the average whilst the grey lines are the individual spectra from structures containing exclusively the motif in question (solid) or mixed motif types (dashed). }
  \label{fig:motif_XAS}
\end{figure}

Finally, we comment on the wide variety of local structure and diverse chemistry observed in the predicted structures, ranging from complex Ge network, to 2D layered Zintl phases, isolated Zintl ions and unusual host-guest structures. The variation in structure with composition was analysed using SOAP descriptors based on the local environments of the Ge atoms only. As shown in Figure \ref{fig:motif_XAS}, as the Na content increases the average number of bonds to Ge neighbours decreases from a maximum of four in the most Ge-rich structures to zero in very Na-rich structures. Additionally, this analysis revealed that triangular \ch{Ge3} clusters are surprisingly abundant within the predicted low-energy Na-rich structures, despite, to the best of our knowledge, not being a known Ge Zintl ion. However, the triangular \ch{Ge3^-} cluster was examined in ref. \cite{tai2011stochastic} and predicted to be only marginally metastable in vacuum. In combination with our predictions, this suggests that triangular \ch{Ge3^-} may be realisable under different conditions or in other chemistries. Inspired by ref. \cite{aarva2021x} an attempt was also made to correlate the the local Ge structure with predicted \ac{XAS} for the Ge K-edge. This analysis identified some distinctive features in the Ge K-edge XAS spectra, shown in Figure \ref{fig:motif_XAS}, and revealed a qualitative relationship between the Ge K-edge position and the local Ge structure as shown in Figure \ref{fig:core-loss}. We anticipate that these results could be useful to those studying non-crystalline materials, with full details given in Section \ref{sec:coreloss}.

The computed phonon and electronic band-structures can be examined interactively using the analysis notebook \cite{zenodo} and confirmed that all structures are dynamically stable and non-magnetic. Simulated \ac{PXRD} patterns and pair-distribution functions are also shown in Figure \ref{fig:pxrd}.

\subsection{10GPa}

\begin{figure*}[t!]
  \includegraphics[width=\textwidth]{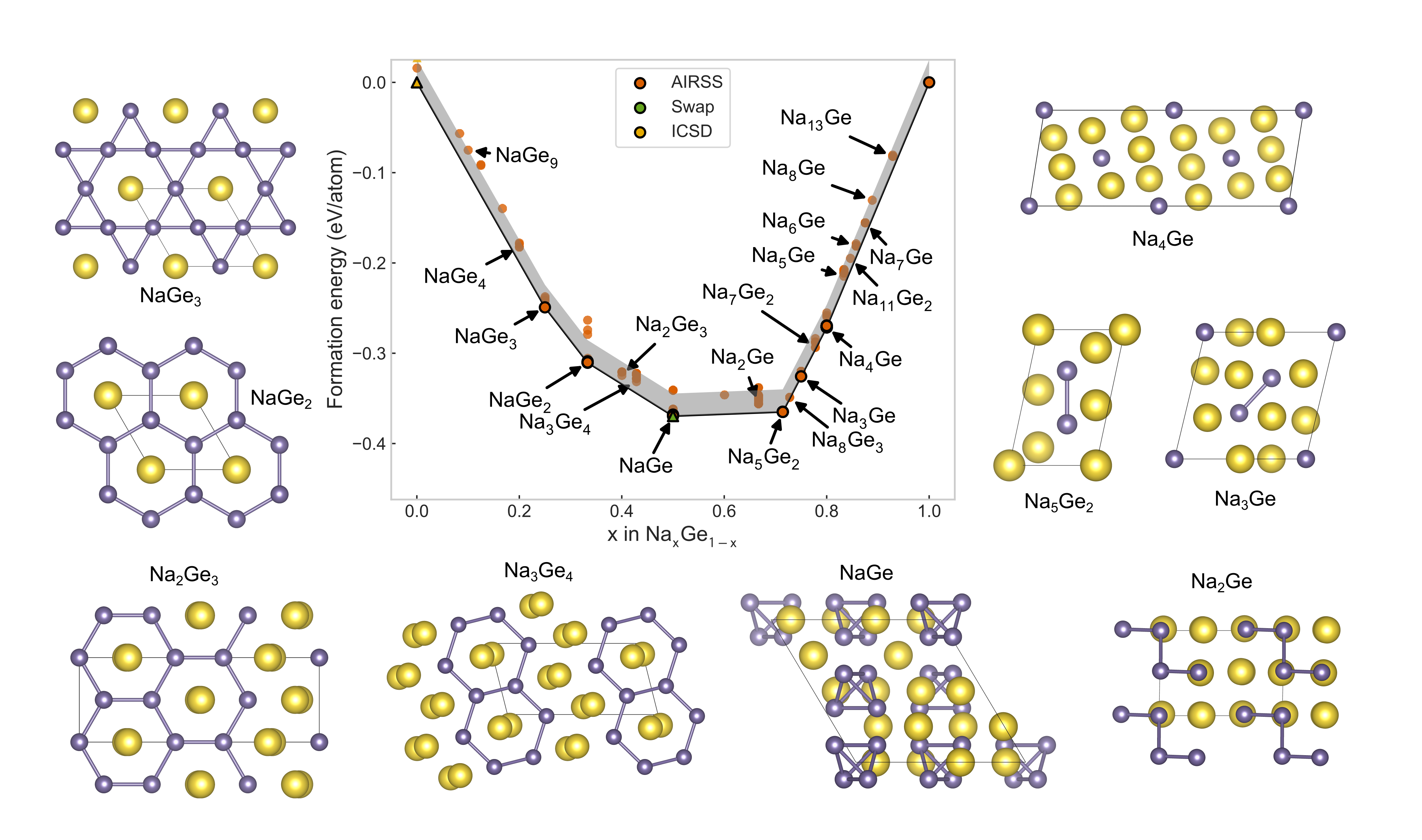}
  \caption{The computed convex hull for the NaGe binary system at 10 GPa. Ge has the $\beta$-Sn structure whilst Na is bcc. The grey band indicates +25 meV/atom. Ge-Ge bonds are drawn between atoms separated by 2.8 \AA~or less.  }
  \label{fig:10GPa_Hull}
\end{figure*}

\begin{table}[h!]
	\centering
	\caption{Na-Ge phases within 25 meV/atom of the 10~GPa convex hull are listed. Compositions calculated to lie on the convex hull are shown in \textbf{bold}. Ge has the $\beta$-Sn structure whilst Na is bcc.  Space groups computed using spglib with symmetry tolerance of 0.01 and packings detected using Ovito \cite{ovito}}
	\begin{tabular}{ccccc}
            \toprule
		formula & $\Delta$ E (meV/atom) & Z & space group & provenance\\
            \midrule
		\textbf{$\beta$-Ge} & - &  2 & I4$_1$/amd & ICSD 53643 \cite{qadri1983high}  \\
            Ge  & 16 &  3 & Cmmm & AIRSS  \\ 
            \ch{NaGe9} & 25 &  1 & P$\bar{1}$ & AIRSS \\
            \ch{NaGe4} & 17 &  2 & P2/m & AIRSS \\
		\textbf{\ch{NaGe3}} & - &  4 & Pna2$_1$ & AIRSS \\
            \textbf{\ch{NaGe2}} & - &  2 & C2/m & AIRSS \\
            \ch{Na2Ge3} & 10 &  2 & Pmma & AIRSS \\
            \ch{Na3Ge4} & 13 &  2 & Cmcm & AIRSS \\
            \textbf{\ch{NaGe}} & - &  16 & I4$_1$/acd &  Swap 409434 \cite{grin1999redetermination} \\
            \ch{Na3Ge2} & 22 &  2 & Cmmm &  AIRSS\\
            \ch{Na2Ge} & 10 &  4 & P$\bar{1}$ &  AIRSS \\%(bcc)\\
            \textbf{\ch{Na5Ge2}} & - &  1 & R$\bar{3}$m & AIRSS \\%(bcc)\\
            \ch{Na8Ge3} & 2 &  1 & R$\bar{3}$m & AIRSS \\%(bcc)\\
            \ch{Na3Ge} & 0.3 &  3 & P2/m & AIRSS \\%(bcc)\\
            \ch{Na7Ge2} & 2 &  1 & P$\bar{3}$m1 &  AIRSS \\%(bcc)\\
            \textbf{\ch{Na4Ge}} & - &  4 & R$\bar{3}$m & AIRSS  \\%(bcc)\\
            \ch{Na5Ge} & 12 &  4 & P$\bar{1}$ & AIRSS \\%(bcc) \\
            \ch{Na11Ge2} & 14 &  1 &  P$\bar{1}$ &  AIRSS \\%(fcc) \\
            \ch{Na6Ge} & 13 &  2 &  P$\bar{1}$ &  AIRSS \\%(fcc) \\
            \ch{Na7Ge} & 14 &  1 & C2/m & AIRSS \\%(bcc) \\
            \ch{Na8Ge} & 20 &  2 & P2$_1$/m &  AIRSS \\%(hcp) \\
            \ch{Na13Ge} & 15 &  1 & P$\bar{6}$ &  AIRSS \\%(hcp) \\
		\textbf{Na} & - & 1 & Im$\bar{3}$m & ICSD 44757 \cite{barrett1956x} \\
  \bottomrule
	\end{tabular}	
	\label{table:NaGe_10GPa}
\end{table}

The computed convex hull at 10 GPa is shown alongside selected low energy structures in Figure \ref{fig:10GPa_Hull}, with further details on structure and provenance provided in Table \ref{table:NaGe_10GPa}. A noteworthy feature of the hull is that the tie-lines connecting the Na-rich phases from Na-\ch{Na5Ge2} form a straight line; indicating a constant formation energy per element within this region. Further examination reveals that all structures on, or close to, the hull from \ch{Na5Ge2}-\ch{Na4Ge} have bcc packing whilst from \ch{Na5Ge}-\ch{Na} a mixture of bcc, fcc and hcp packings are seen.  Within these structures various local orderings of the Ge atoms are found, with Ge-Ge dumbbells (one Ge per lattice site) observed in \ch{Na5Ge2}, \ch{Na8Ge3} and \ch{Na3Ge}, isolated atoms in \ch{Na4Ge}-Na and even \ch{Ge4} chains within \ch{Na2Ge}. Furthermore, no bandgap is seen in the electronic bandstructure for any of these structures. Taken in combination, these observations are consistent with the existence of a Na-rich (likely bcc) metallic solid solution where there is a degree of covalent bonding between any adjacent Ge atoms; see ELF isosurfaces of \ch{Na2Ge}, \ch{Na5Ge2} and \ch{Na3Ge} in Figure \ref{sfig:ELF_10GPa_Na-rich}.

\begin{figure}[h!]
  \includegraphics[width=0.48\textwidth]{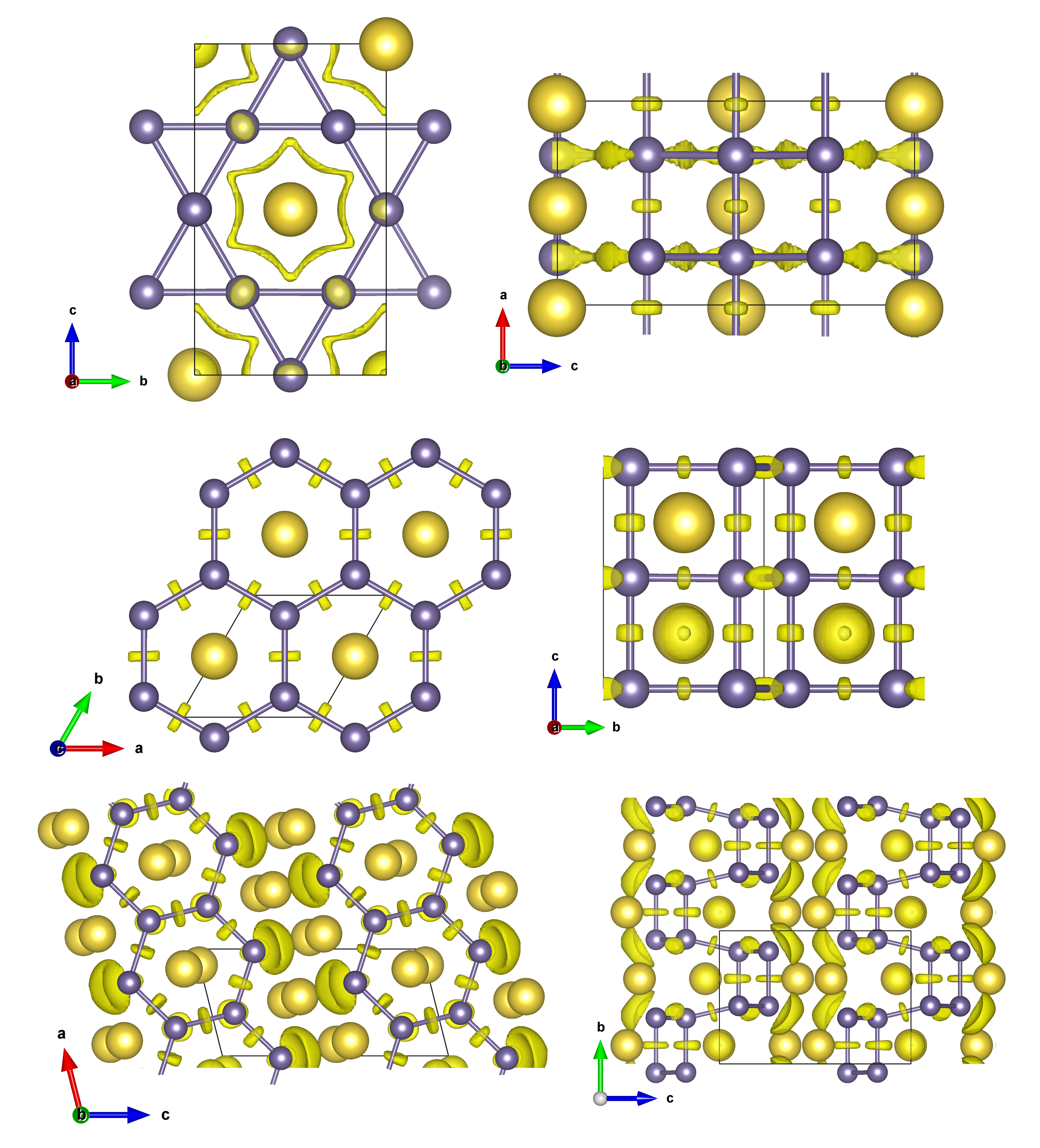}
  \caption{ELF isosurfaces are shown top-to-bottom for \ch{NaGe3}, \ch{NaGe2} and \ch{Na3Ge4}. The isosurface level is 0.57 for the top two structures and 0.6 for \ch{Na3Ge4}.}
  \label{fig:10GPa_ELF}
\end{figure}

Unsurprisingly, the highly stable NaGe structure, an I4$_1$/acd packing of Na atoms and \ch{Ge4} tetrahedra, remains stable at 10 GPa. However, despite this similarity to the 0 GPa results, the low energy Ge-rich structures are considerably different, with significantly less variety seen. In fact, all low energy Ge-rich structures can be viewed as derivatives of the high pressure simple-hexagonal phase referred to as Ge-V \cite{mujica2001high, kelsall2021high}. 
Ge-V consists of hexagonally close packed layers in an AA stacking sequence. The intra-layer bonding is delocalised whilst the inter-layer bonds have covalent character, as visible in \ref{sfig:ELF_10GPa_sh-Ge}. 
The predicted \ch{NaGe3} and \ch{NaGe2} structures are both related to Ge-V by deleting columns of Ge atoms (perpendicular to the layers) and replacing them with Na atoms, shifted by half the layer spacing so that the Na atoms sit midday between the Ge layers. 
Replacing different columns gives rise to the graphite-like structure present in \ch{NaGe2} and the kagome structure found in \ch{NaGe3}, as illustrated in \ref{sfig:sh-Ge_schematic}. 
The other low energy structures near the hull arise either from replacing fewer, \ch{NaGe9}, \ch{NaGe7} and \ch{NaGe4}, or more, \ch{Na2Ge3} and \ch{Na3Ge4}, columns of Ge atoms in the same way. 
Note that both \ch{Na2Ge3} and \ch{Na3Ge4} contain ``buckled'' Ge-hexagons which arose from following phonon modes with imaginary frequencies. 
The computed \ac{ELF} shown in Figure \ref{fig:10GPa_ELF} suggests that this buckling allows the Ge atoms in \ch{Na3Ge4} to adopt a near tetrahedral geometry with 3 bonds to adjacent Ge atoms and a single lone pair. 
In contrast to this, both \ch{NaGe3} and \ch{NaGe2} have flat layers, with the Ge within \ch{NaGe2} adopting an unusual 5-fold coordinated trigonal bipyramidal geometry. 

\begin{figure}[h!]
  \includegraphics[width=0.48\textwidth, trim={0 0 0 1.5cm}, clip]{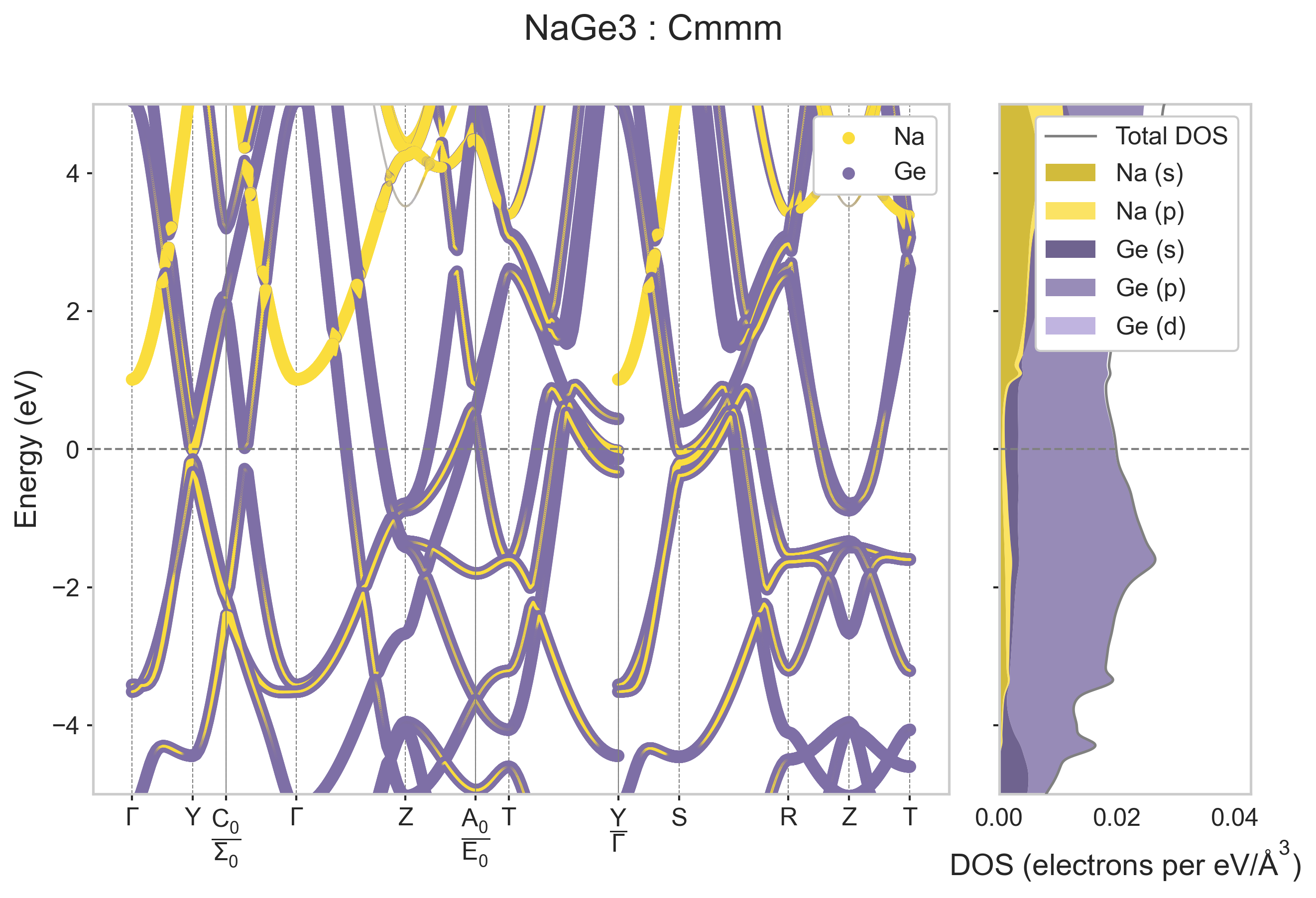}
  \caption{Electronic bandstructure of Cmmm \ch{NaGe3} at 0 GPa projected onto the Na and Ge states and the corresponding density of states projected onto the Na $s$ and $p$ and Ge $s$, $p$ and $d$ states. \texttt{OptaDOS} \cite{morris2014optados} was used for the projection of states. }
  \label{fig:0GPa_NaGe3_bs}
\end{figure}

The localised intra-layer and delocalised inter-layer bonding seen in simple hexagonal Ge-V is retained in \ch{NaGe3} and is almost identical to that seen in the recently predicted \ch{MgSi3} \cite{zha2023refined}. The exceptionally diverse electronic physics \cite{hu2023electronic} associated with the kagome topology has inspired enormous interest in recent years, with many such materials, including the isostructural \ch{MgSi3} and \ch{MgB3} \cite{an2023topological} found to be superconducting at low temperatures. A full investigation into the electronic properties of \ch{NaGe3} is beyond the scope of this work, however, we note that \ch{NaGe3} is both dynamically stable at 0 GPa and that the electronic bandstructure, see Figure \ref{fig:0GPa_NaGe3_bs} is markedly similar to that reported for \ch{MgSi3} \cite{zha2023refined}; crucially the dominant contribution to the electronic density of states near the Fermi level is from the Ge $p$ orbitals, suggesting that the electronic properties will be largely determined by the kagome lattice. Coupling these observations with the bonding similarity observed in the computed \acp{ELF} makes it plausible that \ch{NaGe3} is also a superconductor and merits further investigation.

As mentioned previously, the main aim of the 10 GPa searches was to discover structures with covalent Ge-frameworks and investigate the structural changes upon removing the Na at 0 GPa. Whilst \ch{NaGe3} remains dynamically stable at 0 GPa, after removing the Na the Ge lattice is unstable and transforms into the simple hexagonal Ge-V during structural relaxation. In contrast, \ch{NaGe2} transforms into a distorted hexagonal lonsdalite Ge framework containing Na. As expected, after removing the Na the structure transforms into pure lonsdalite Ge. The stability of both \ch{NaGe2} and lonsdalite Ge hints that intermediate phases, with variable Na content might also be synthesisable and that the optical absorption could be tuned in this way.

Finally, we note that both non-polarised and spin-polarised electronic bandstructures were computed for all low-energy structures. These can be viewed using the analysis notebook \cite{zenodo} and confirm that all materials are predicted to be non-magnetic. As at 0 GPa, the temperature dependent convex hull was computed using the harmonic approximation and is provided in Figure \ref{sfig:Thull_10GPa}, where it is clear that the 10 GPa hull remains essentially unchanged at temperatures below 1000 K. 

\section{Conclusions} \label{sec:conclusion}
In this study, a combination of element swapping from known phases, AIRSS and the \texttt{illustrado} genetic algorithm were used to search for stable binary phases of sodium-germanide at both 0 and 10 GPa. The searches at 0 GPa proved exceptionally challenging, which we attribute to the diversity in low energy local Ge structure, so a \ac{MLIP} was trained and used to carry out over one million additional structural relaxations. Phonon calculations were used to assess the dynamical stability of predicted structures and to compute the temperature dependent vibrational contribution the free energy.

At 0 GPa all experimentally known structures Na$_\delta$Ge$_{34}$, \ch{Na4Ge13}, \ch{Na12Ge17} and \ch{NaGe} were found to lie on, or close to, the predicted convex hull. Additionally, we predict three new phases, \ch{Na3Ge2}, \ch{Na2Ge} and \ch{Na4Ge9} to be on the convex hull within \ac{DFT} error. These phases are all Zintl phases with \ch{Na2Ge} containing tetrahedral \ch{Ge4} ions, as found in \ch{NaGe}, whilst \ch{Na2Ge} and \ch{Na4Ge9} both contain negatively charged \ch{Ge2} dumbbells. In principle, the stability of \ch{Na9Ge4} increases the theoretical capacity of the germanium anode from 369~mAhg$^{-1}$, based on \ch{NaGe} as the endpoint, to 830~mAhg$^{-1}$. However, we note that as Na cannot be intercalated into crystalline Ge it remains to be seen if this increase is practicaly useful.

The high pressure searches confirmed that I$4_1$/acd packing of germanium tetrahedra and Na atoms remains stable at 10 GPa and suggest the existence of a bcc solid solution between \ch{Na5Ge2} and Na metal. Two stable Ge-rich structures with covalently bonded Ge frameworks are predicted to be stable. \ch{NaGe3} contains kagome layers and is isostrutural with \ch{MgB3} and \ch{MgSi}, both of which are superconducting at low temperature, suggesting that the electronic structure of \ch{NaGe3} merits further investigation. \ch{NaGe2} consists of a simple hexagonal Ge lattice with Na atoms filling the hexagonal channels. This structure transforms into Na-filled lonsdalite Ge at 10 GPa and then pure lonsdalite Ge if the Na is fully removed. Varying the level of Na within the lonsdalite structure may provide a mechanism to tune the optical properties of lonsdalite Ge.

\section{Acknowledgements} \label{sec:acknowledgements}
The authors would like to thank Phoebe K. Allan, Mathew L. Evans,  Andrew L. Goodwin and Bartomeu Monserrat for helpful discussions. JPD acknowledges the financial support of a Sims Scholarship for PhD funding.
AJM acknowledges support from EPSRC via CCP-NC (EP/T026642/1), CCP9 (EP/T026375/1), UKCP (EP/P022561/1) and Baskeville (EP/T022221/1).
AFH acknowledges the financial support of the Gates Cambridge Trust and the Winton Programme for the Physics of Sustainability, University of Cambridge. 
This work was performed using resources provided by the Cambridge Service for Data Driven Discovery (CSD3) operated by the University of Cambridge Research Computing Service (www.csd3.cam.ac.uk), provided by Dell EMC and Intel using Tier-2 funding from the Engineering and Physical Sciences Research Council (capital grant EP/T022159/1), and DiRAC funding from the Science and Technology Facilities Council (www.dirac.ac.uk).
We are grateful for computational support from the UK national high performance computing service, ARCHER2, for which access was obtained via the UKCP consortium and funded by EPSRC grant ref EP/X035891/1.

\bibliographystyle{PRL}
\bibliography{refs}

\clearpage
\widetext
\begin{center}
\section{Supporting Information}
\end{center}
\setcounter{figure}{0}
\renewcommand{\theequation}{S\arabic{equation}}
\renewcommand{\thefigure}{S\arabic{figure}}
\renewcommand{\theHfigure}{Supplement.\thefigure}

\subsection{NNP}
\begin{figure}[h]
     \centering
     \begin{subfigure}[b]{0.47\textwidth}
         \centering
         \includegraphics[width=\textwidth]{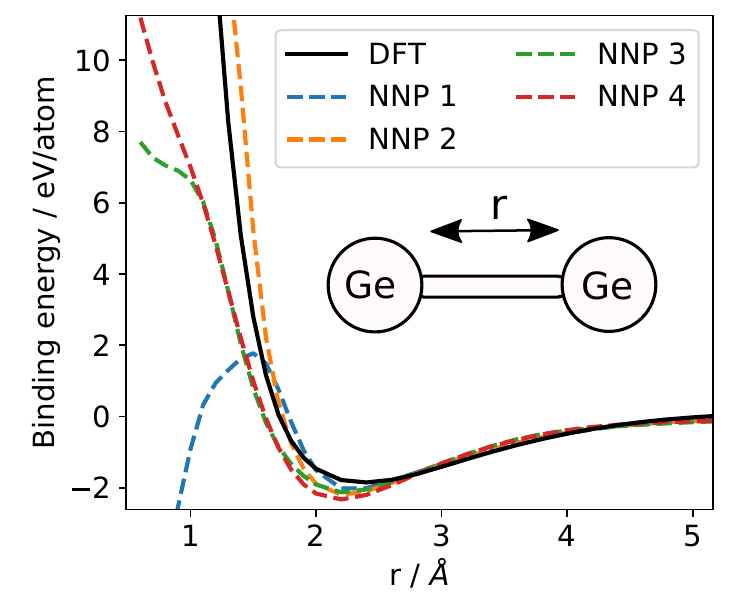}
         \caption{}
         \label{fig:dimer_ensemble}
     \end{subfigure}
     \hfill
     \begin{subfigure}[b]{0.5\textwidth}
         \centering
         \includegraphics[width=\textwidth]{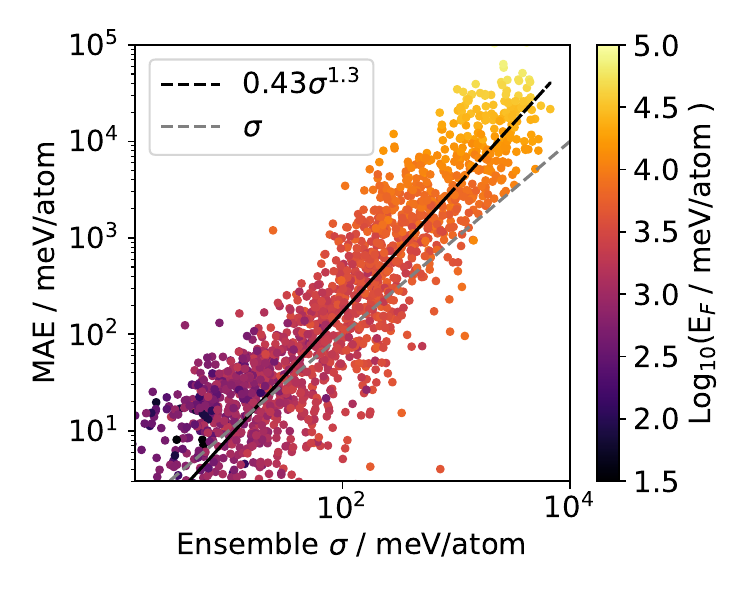}
         \caption{ }
         \label{fig:ensemble_10GPa}
     \end{subfigure}
     \caption{a) The predicted energy of the Ge-Ge dimer is shown for the four members of the initial \ac{MLIP} ensemble. The committee variance is very large at short distances as there is no training data in this region. b) The mean absolute error (MAE) in the energy predicted by the \ac{MLIP} ensemble is shown against the standard deviation of the ensemble for a randomly selected set of snapshots taken from the \ac{AIRSS} at 10 GPa. }
\end{figure}

\begin{figure}[h]
     \centering
     \begin{subfigure}[b]{0.47\textwidth}
         \centering
         \includegraphics[width=\textwidth]{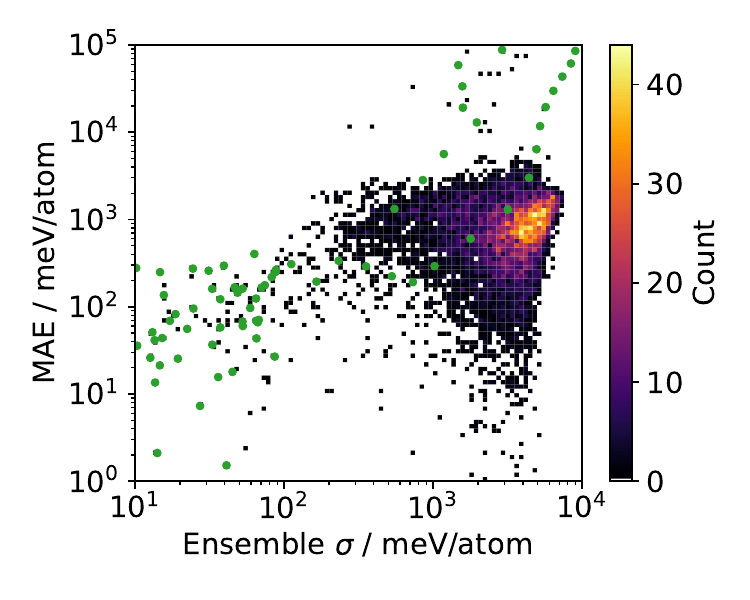}
         \caption{  }
         \label{fig:plus_heatmap}
     \end{subfigure}
     \hfill
     \begin{subfigure}[b]{0.47\textwidth}
         \centering
         \includegraphics[width=\textwidth]{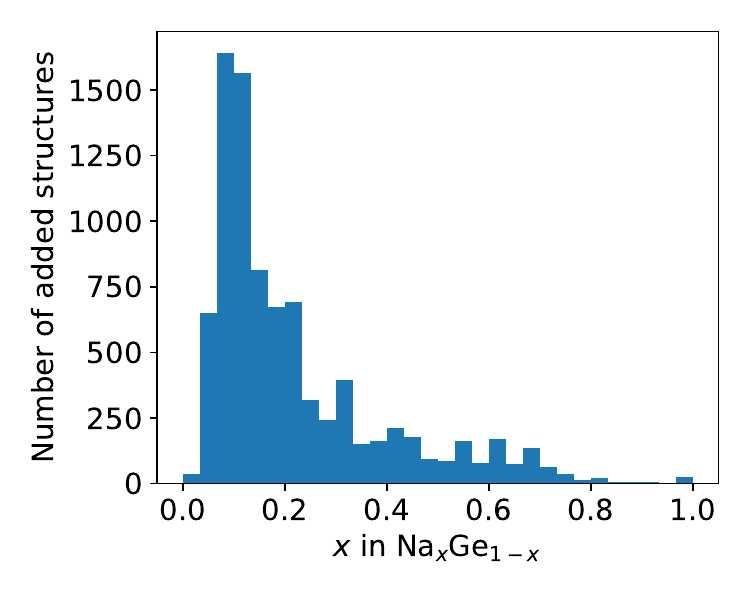}
         \caption{ }
         \label{fig:plus_stoich}
     \end{subfigure}
    \caption{a) The mean absolute error (MAE) in the energy predicted by the \ac{MLIP} ensemble is shown against the standard deviation of the ensemble for the $\sim8700$ structures added to the training set. Dimers are indicated in green. Note that the average MAE on these structures is $\sim$1 eV/atom compared to only 15 meV/atom on the test set. b) The distribution of stoichiometries for the structures added to the training set using the standard deviation in the \ac{MLIP} committee as the selection criterion.}
\end{figure}

\newpage
The final \texttt{aenet} potential used a Chebyshev type basis with 12 radial and 12 angular functions. The radial cutoff was 8~\AA~ whilst the angular cutoff was 4.5~\AA. The chemical potentials used were $E_\mathrm{Na} = -1304.21509$ eV and $E_\mathrm{Ge} = -107.28931325$ eV, corresponding to the energies of bcc Na and diamond Ge respectively. A separate potential was trained for each central element. Both potentials used 2 layer neural networks with 15 nodes in each layer and tanh activation functions. The Levenberg-Marquardt optimisation algorithm was used with the full training set as the batch size and parameters of \texttt{learnrate=0.1d0},  \texttt{iter=3}, \texttt{conv=0.001} and \texttt{adjust=5.0}.

We note that version 2.04 of \texttt{aenet} was used, so that only energies were used during training - typically energies, forces and virials are used when training \acp{MLIP}. However, despite this reasonable accuracy was achieved on force predictions. We primarily attribute this to the large amount of reference training data that was used. We also speculate that the small size and symmetric nature of the training structures meant that the energy carried relatively more information (compared to the forces) than if large asymmetric structures were used.

\begin{figure}[h!]
    \centering
    \includegraphics[width=0.95\textwidth]{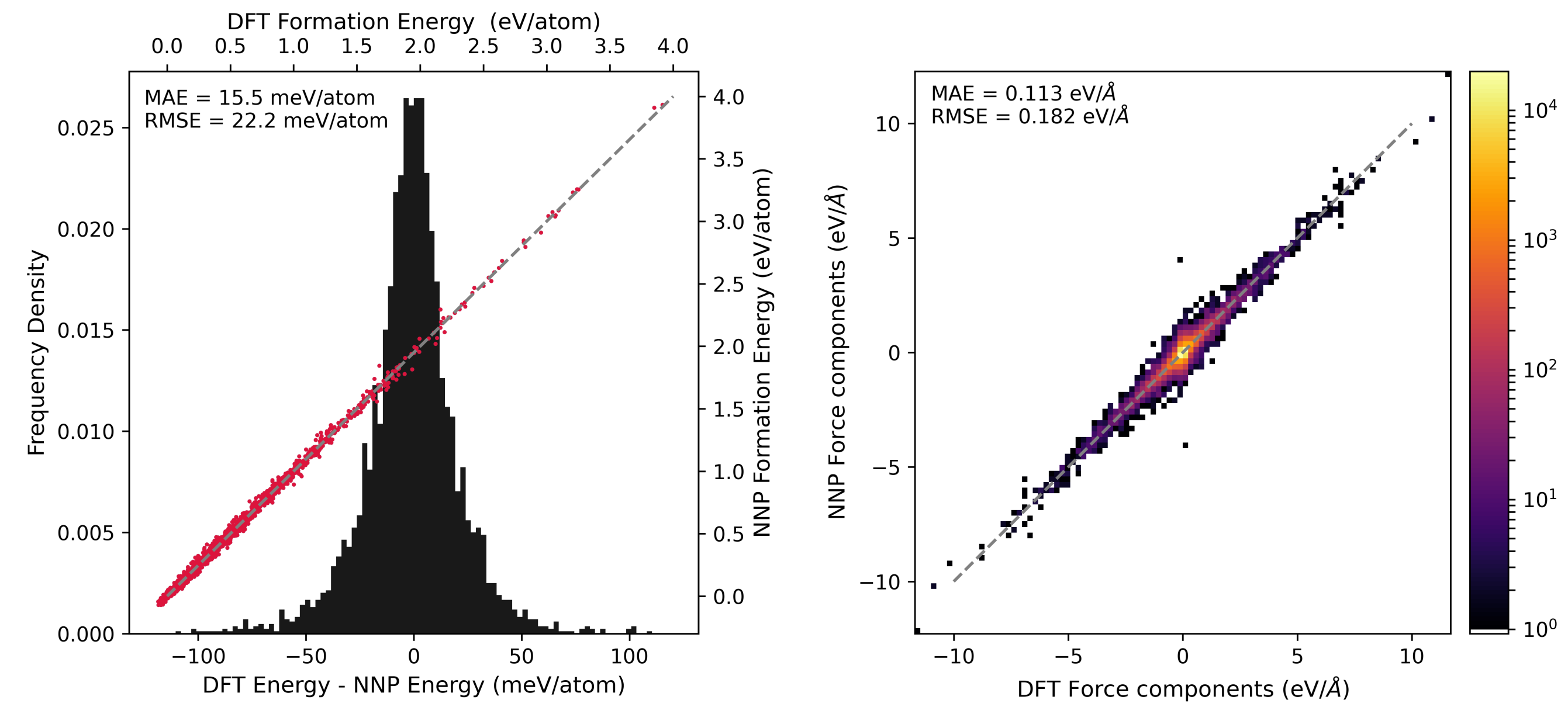}
    \caption{The energies and forces predicted by the \ac{MLIP} on the test set are compared to the \ac{DFT} reference values. Force components which are exactly zero due to symmetry were excluded when computing the errors and are not shown in the figure.}
    \label{fig:NNP_performance}
\end{figure}

\clearpage
\subsection{ELF}

\begin{figure}[h!]
    \centering
    \includegraphics[width=0.8\textwidth]{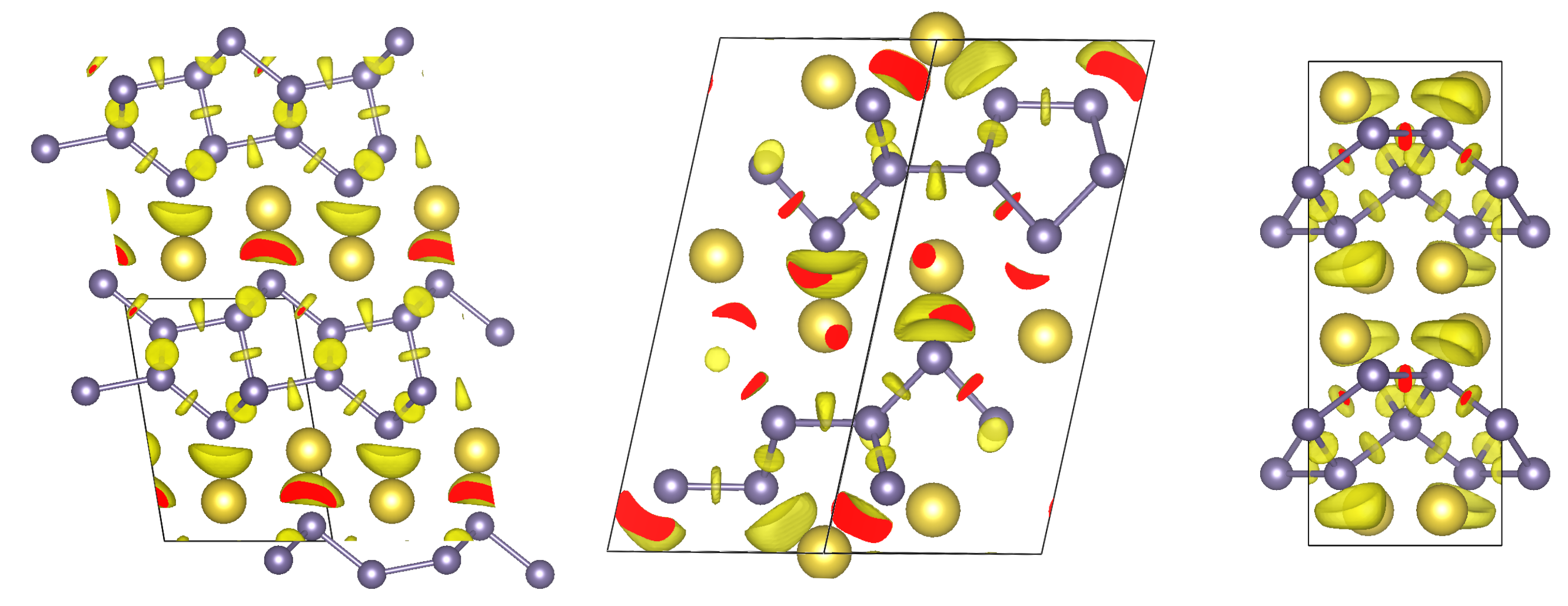}
    \caption{The Electron Localisation Function (valence electrons) is shown for the \ch{NaGe3}, \ch{NaGe2} and \ch{Na2Ge3} phases at 0 GPa. The isosurface value is 0.65. The ratio of 3-fold:4-fold coordinated Ge in \ch{NaGe3}, \ch{NaGe2} and \ch{Na2Ge3} is 1:2, 1:1 and 2:1 respectively. This can be understood as each Na atom donating an electron to one of the Ge atoms, which then behaves as a Group XV element like P with a single lone pair, and so forms three bonds, whilst the remaining Ge atoms form four bonds.}
    \label{sfig:ELF_layered_zintl}
\end{figure}

\begin{figure}[h!]
    \centering
    \includegraphics[width=0.8\textwidth]{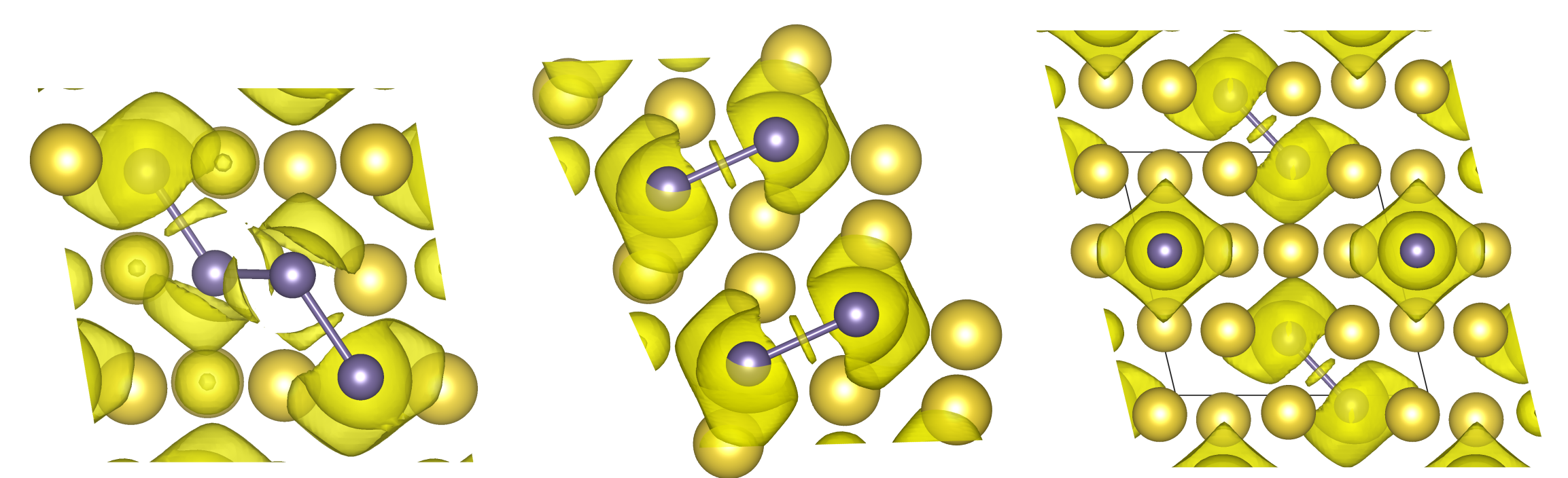}
    \caption{The Electron Localisation Function (valence electrons) is shown for the \ch{Na2Ge}, \ch{Na5Ge2}, and \ch{Na3Ge} phases at 10 GPa. The isosurface value is 0.57. From left to right, tjhe ELFs show \ch{Ge4} chains, \ch{Ge2} dumbbells  and a mixture of \ch{Ge2} dumbbells and isolated atoms.}
    \label{sfig:ELF_10GPa_Na-rich}
\end{figure}

\begin{figure}[h!]
    \centering
    \includegraphics[width=0.4\textwidth]{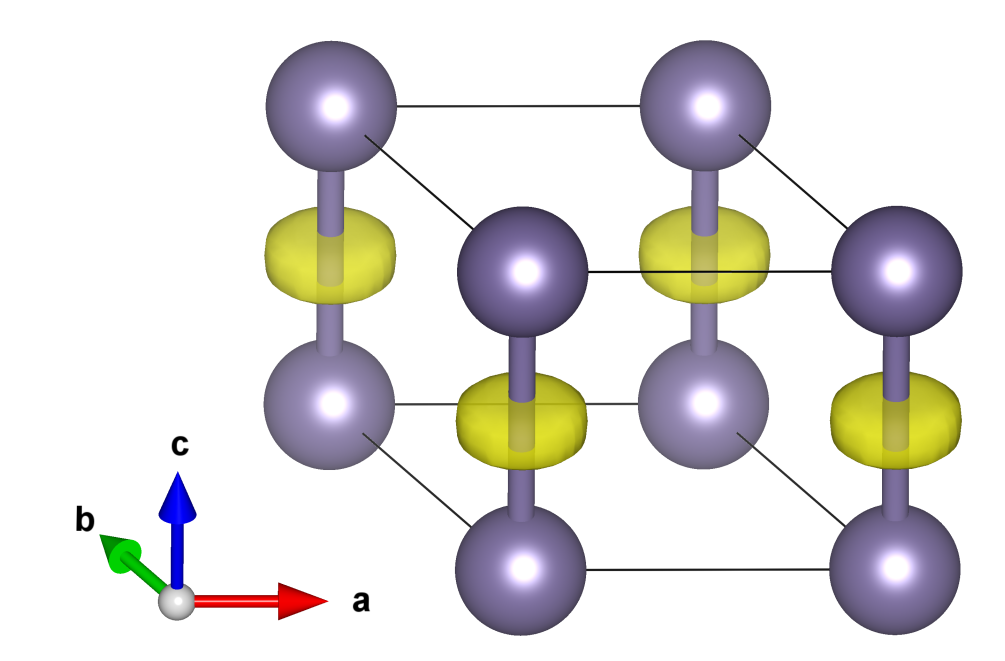}
    \caption{The Electron Localisation Function (valence electrons) is shown for sh Ge at 10 GPa. The isosurface value is 0.57. Covalent bonds are seen between layers but not within the layers themselves.}
    \label{sfig:ELF_10GPa_sh-Ge}
\end{figure}

\clearpage
\subsection{Search Hulls}

\begin{figure}[h!]
    \centering
    \includegraphics[width=0.8\textwidth]{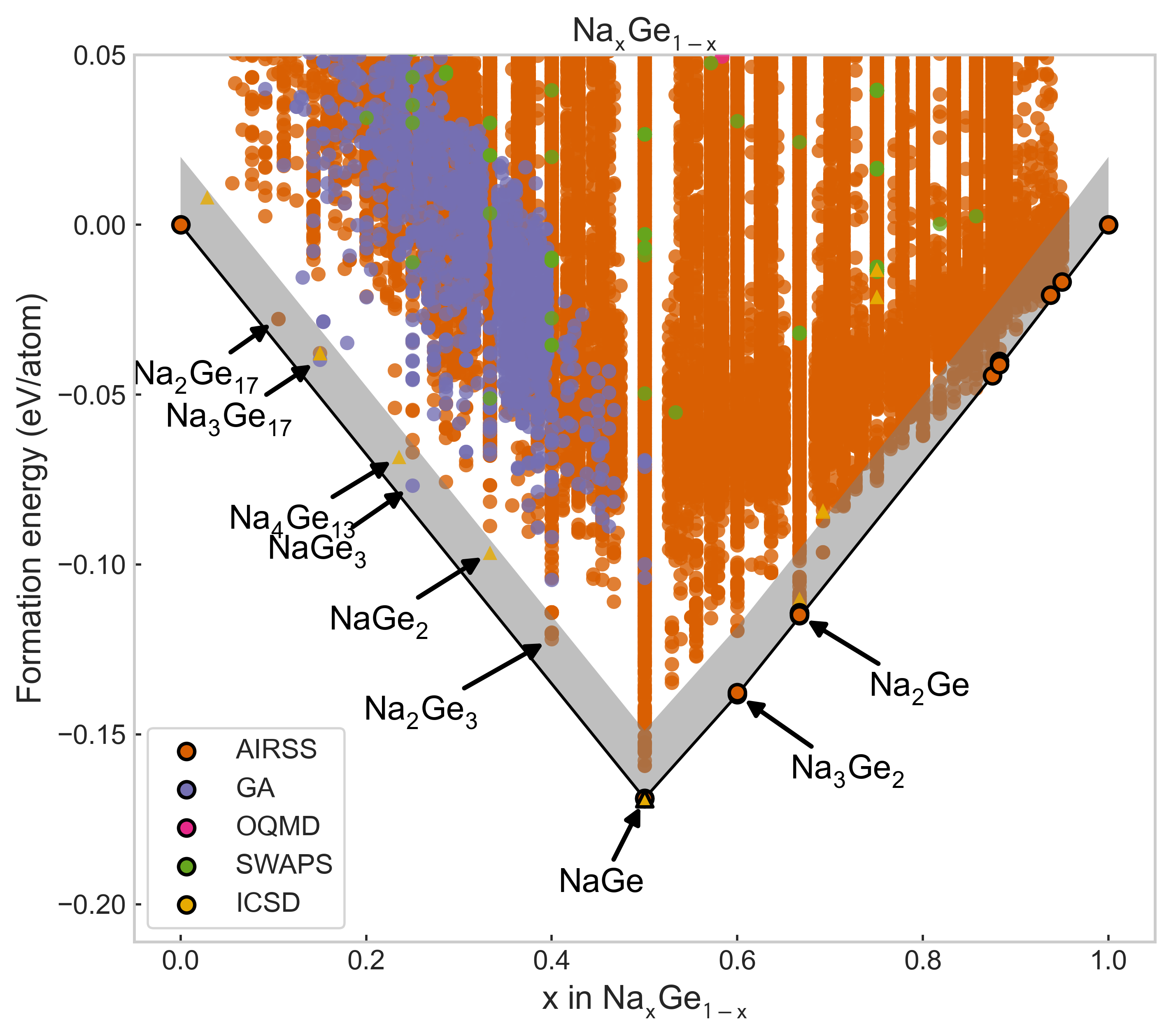}
    \caption{The 0~GPa convex hull computed at ``search accuracy'', 400 eV plane-wave cutoff etc. Note that despite the large number of searches performed, very few Ge-rich structures were found near the hull. }
    \label{sfig:0GPa_search_hull}
\end{figure}

\clearpage
\subsection{Phonons and temperature dependence}

\begin{figure}[h!]
  \includegraphics[width=0.48\textwidth]{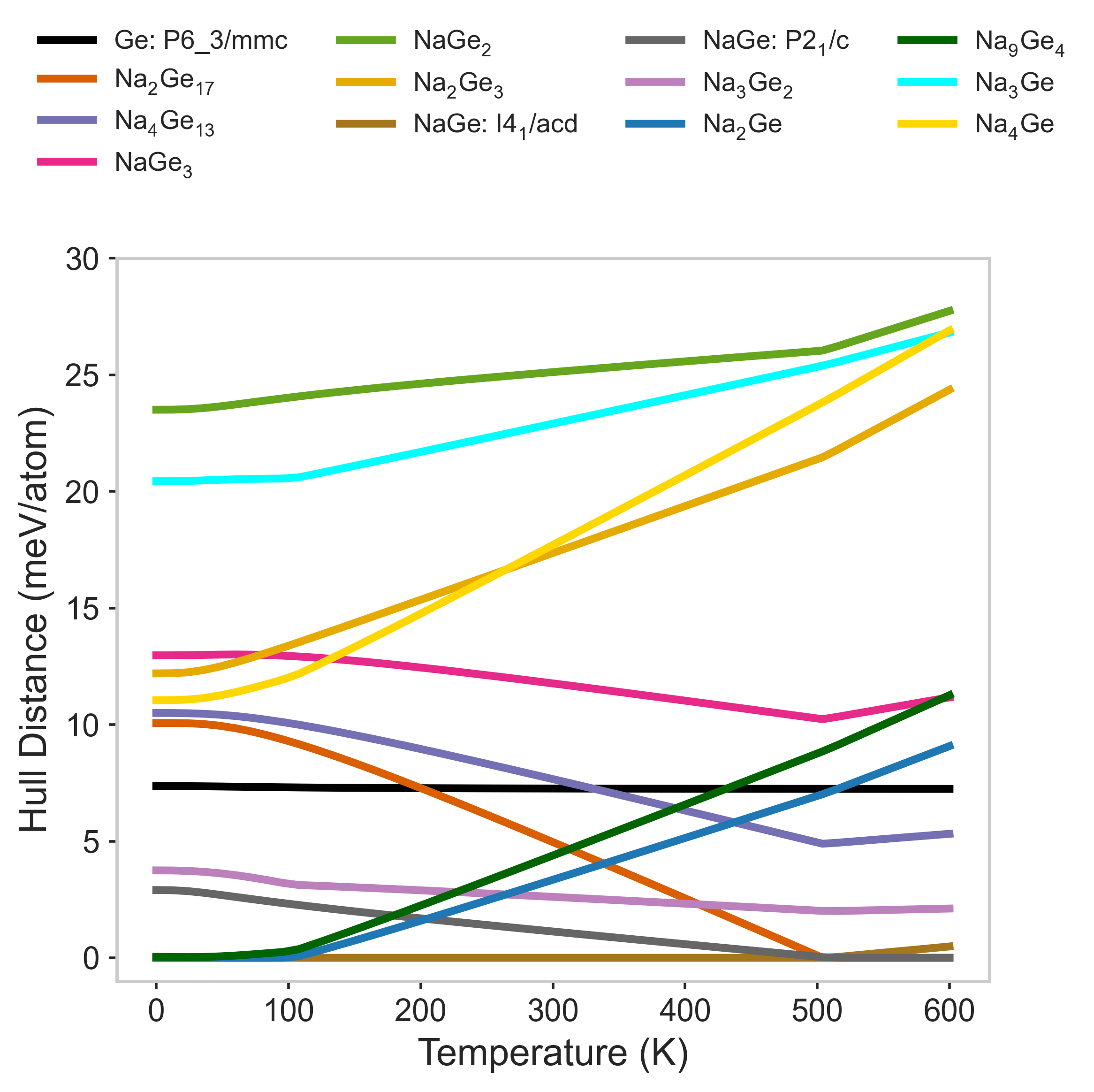}
  \caption{Distances to 0 GPa hull as a function of temperature. The kinks arise where structures move on/off of the convex hull line. The experimentally known \ch{Na2Ge17} and \ch{Na4Ge13} are both relativey stabilised at higher temperatures. }.
  \label{sfig:Thull_dists}
\end{figure}

\begin{figure}[h!]
  \includegraphics[width=0.48\textwidth]{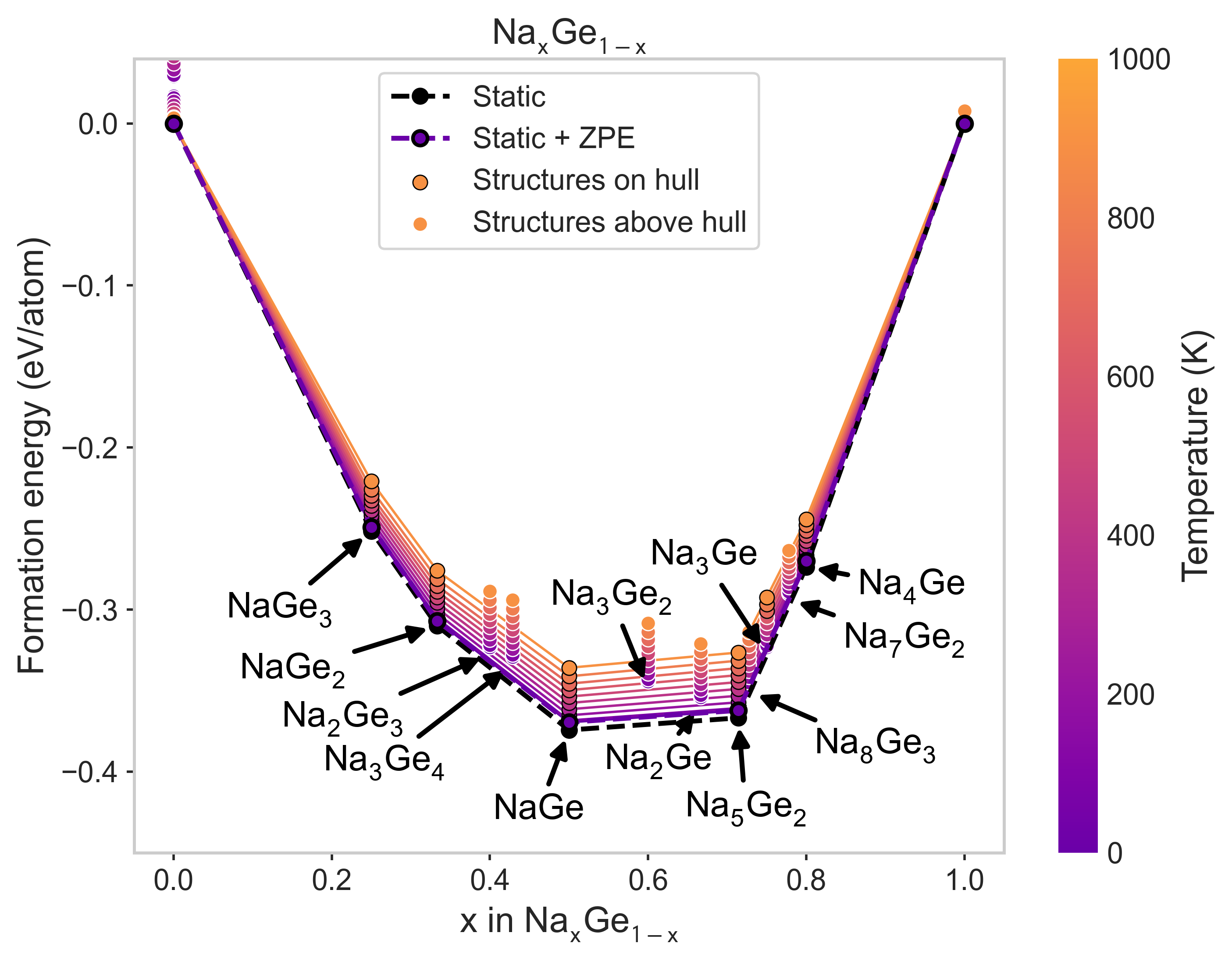}
  \caption{10 GPa temperature dependent convex hull computed using vibrational contributions to the free energy from phonon modes in the harmonic approximation. The shape of the hull remains remarkably constant.}
  \label{sfig:Thull_10GPa}
\end{figure}

\begin{figure}[h!]
  \includegraphics[width=0.48\textwidth]{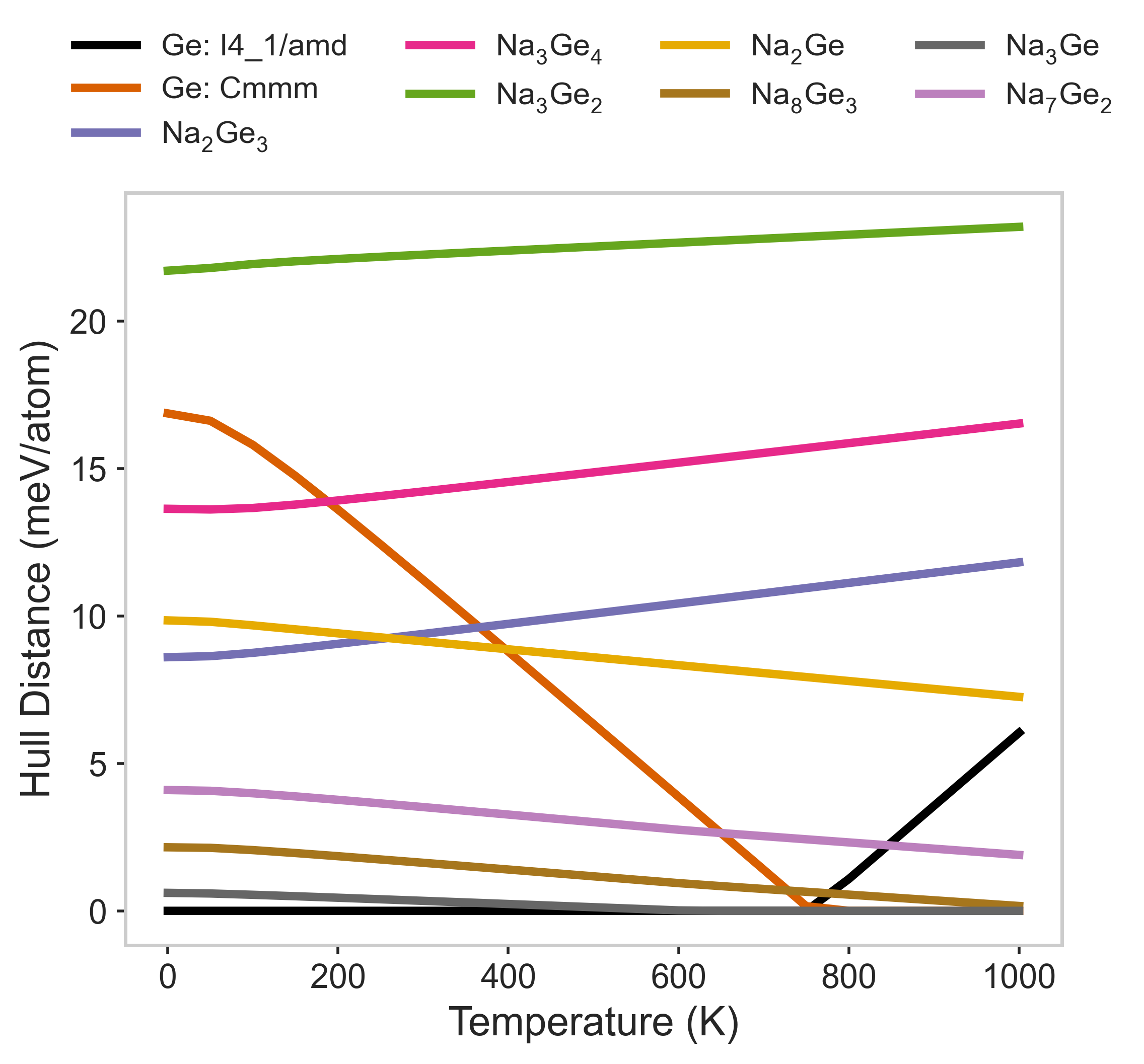}
  \caption{Distances to 10 GPa hull as a function of temperature. The kinks arise where structures move on/off of the convex hull line. Ge: Cmmm is a distorted version of simple hexagonal Ge-V. }.
  \label{sfig:Thull_10_dists}
\end{figure}

\begin{figure}[h!]
  \includegraphics[width=0.48\textwidth]{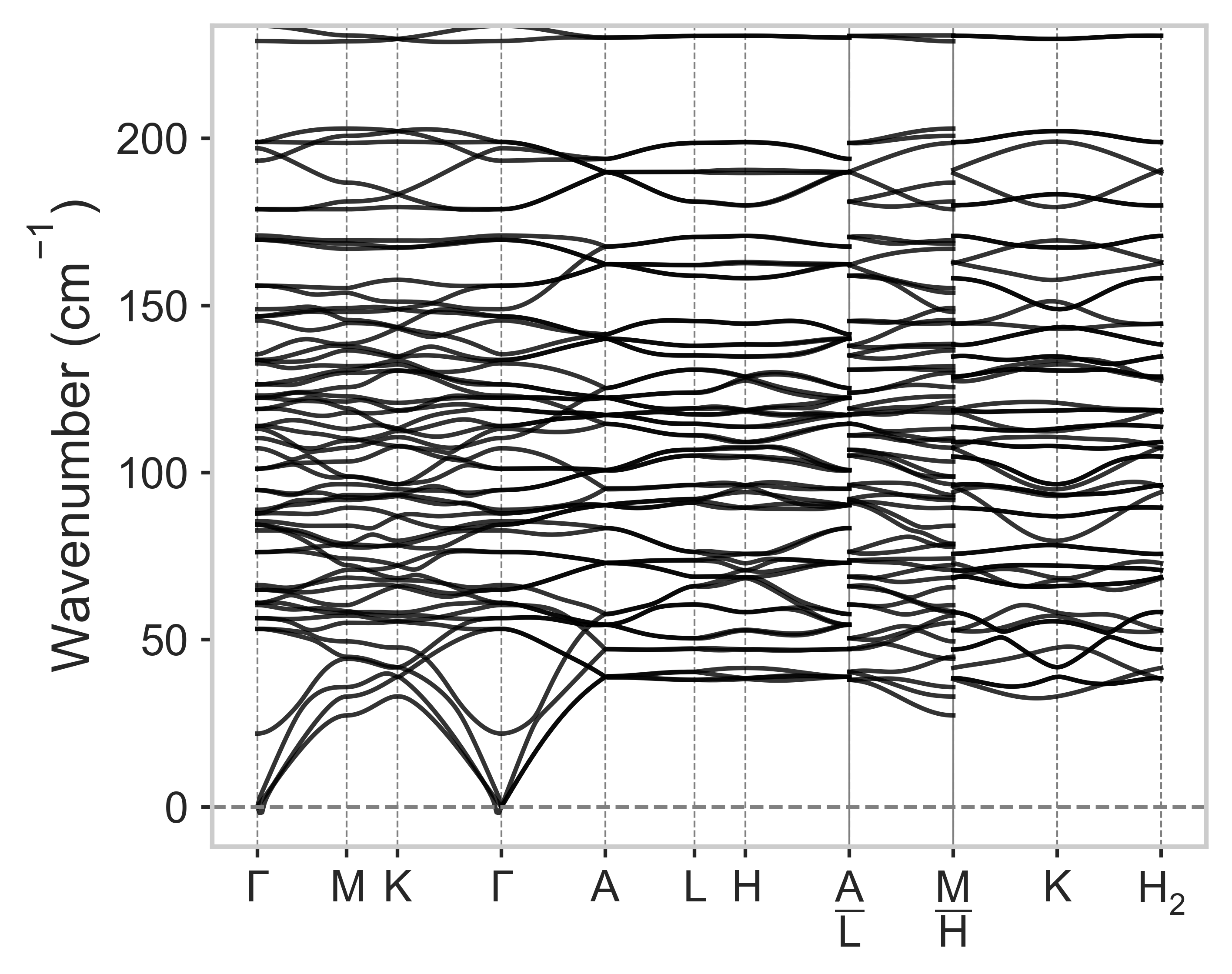}
  \caption{The computed phonon bandstructure of \ch{Na3Ge2}. The very small region of imaginary frequency (shown as negative) near the $\Gamma$ point is an artifact from the Fourier interpolation, as discussed in ref. \cite{gaal2008accelerating}.}
  \label{fig:Na3Ge2_phonon_bs}
\end{figure}

\clearpage
\subsection{Ge-IV compared to \ch{NaGe3} and \ch{NaGe2}}

\begin{figure*}[h!]
  \includegraphics[width=0.65\textwidth]{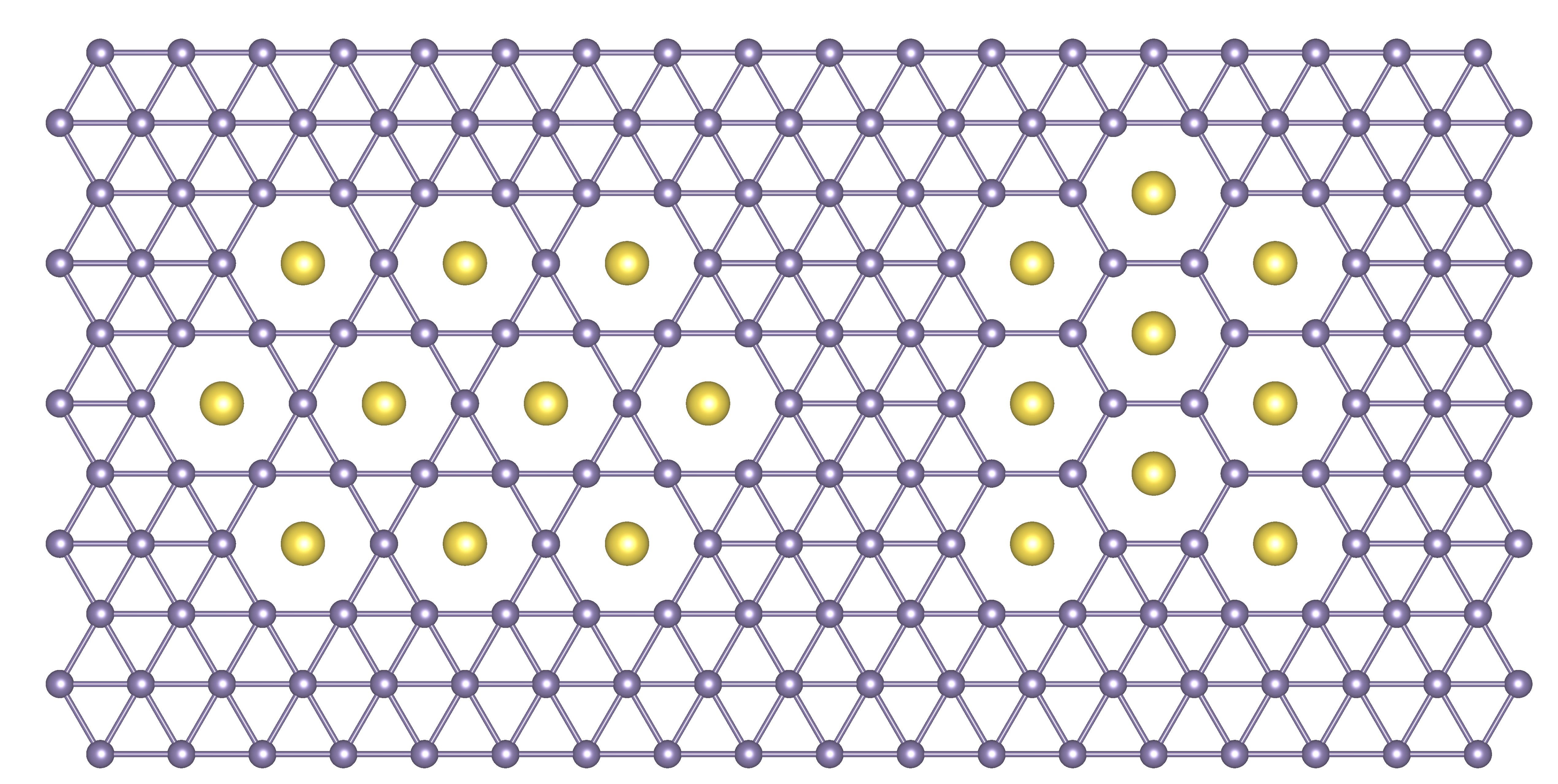}
  \caption{The structural relation between simple-hexagonal Ge (background), \ch{NaGe3} (left) and \ch{NaGe2} (right) is shown. The Na atoms sit halfway between the Ge layers}
  \label{sfig:sh-Ge_schematic}
\end{figure*}

\subsection{Ge$_\delta$\ch{Ge34}}

\begin{figure*}[h!]
  \includegraphics[width=0.65\textwidth]{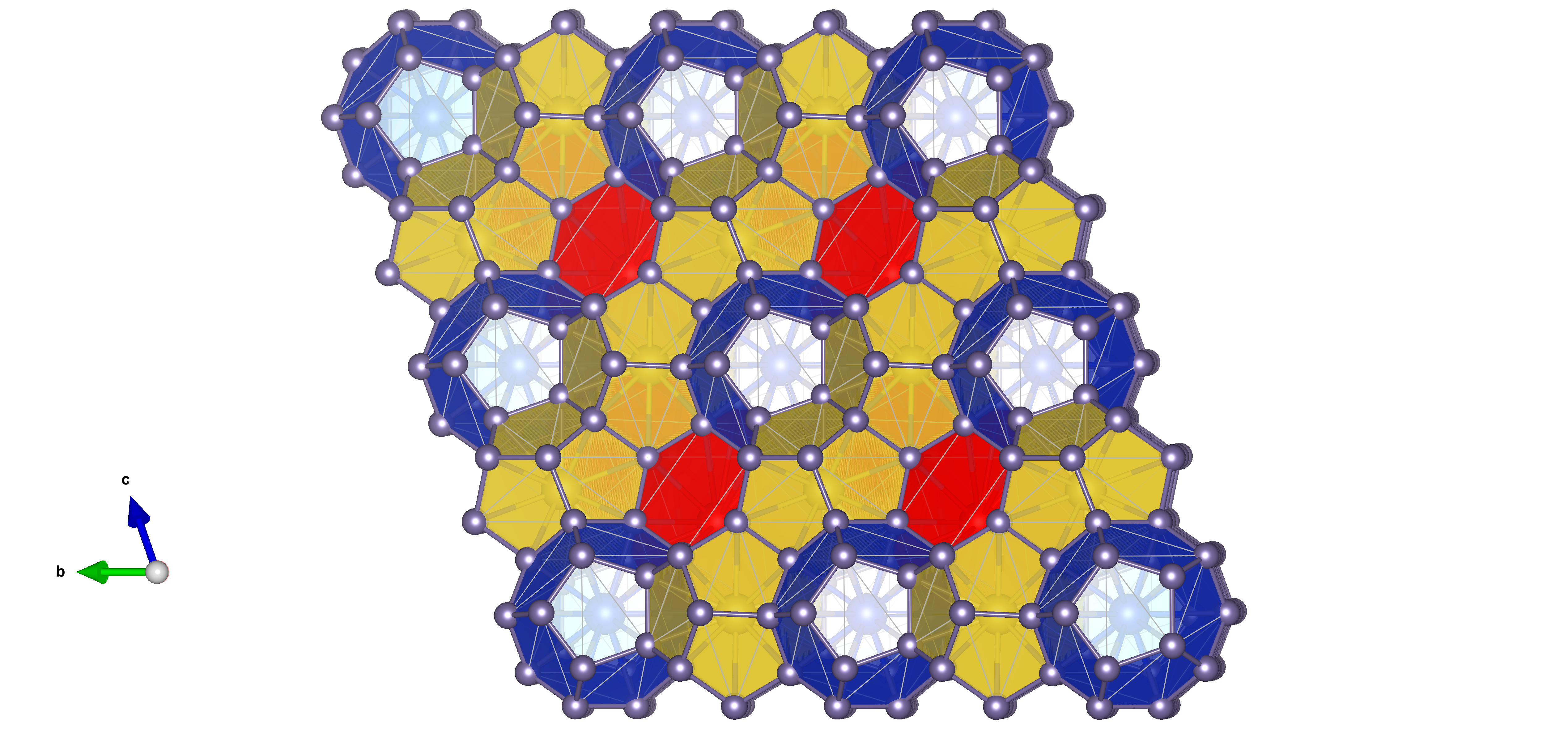}
  \caption{\ch{NaGe34}, \ch{Na2Ge17} and \ch{Na3Ge17} all share the same Ge clathrate structure so are shown as Na$_\delta$\ch{Ge34} with the Na colored blue, blue and yellow and blue, yellow and red for $\delta$=1, 4 and 6 respectively. The yellow and blue Na sit at the centre of the \ch{Ge20} dodecahedra whilst the red Na sit off-centre in the \ch{Ge28} hexakaidecahedra.}
  \label{sfig:Na6Ge34_cages}
\end{figure*}

\clearpage
\subsection{Electronic bandstructures}

\begin{figure}[h!]
    \centering
    \begin{tikzpicture}
      \node[inner sep=0pt,anchor=east] (b) at (0cm, 0cm)
        {\includegraphics[width=9cm,keepaspectratio, trim={0 0 0 0.82cm},clip]
        {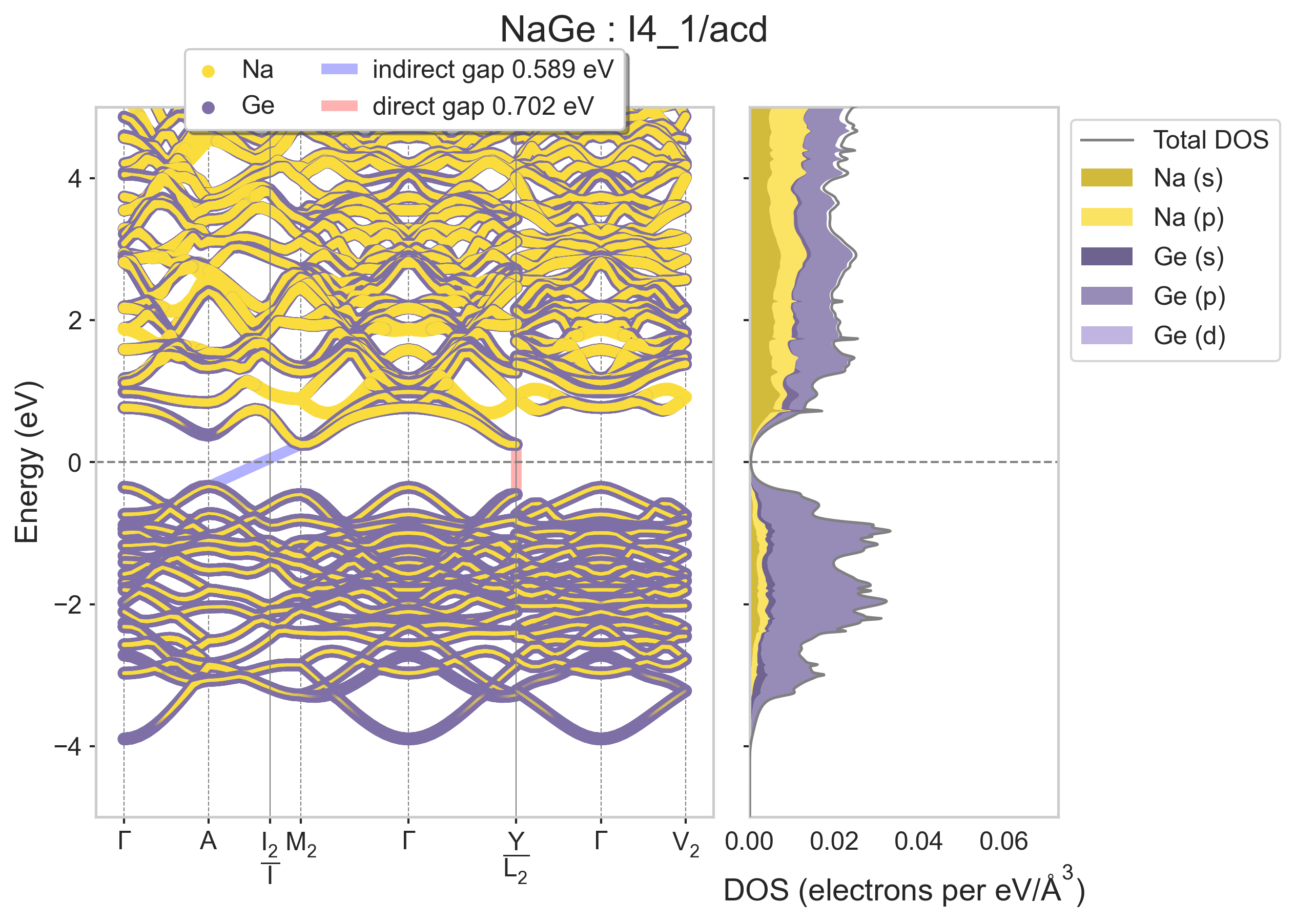}};
       \node[inner sep=0pt,anchor=east] (b) at (9cm, 0cm)
        {\includegraphics[width=9cm,keepaspectratio, trim={0 0 0 0.82cm},clip]
        {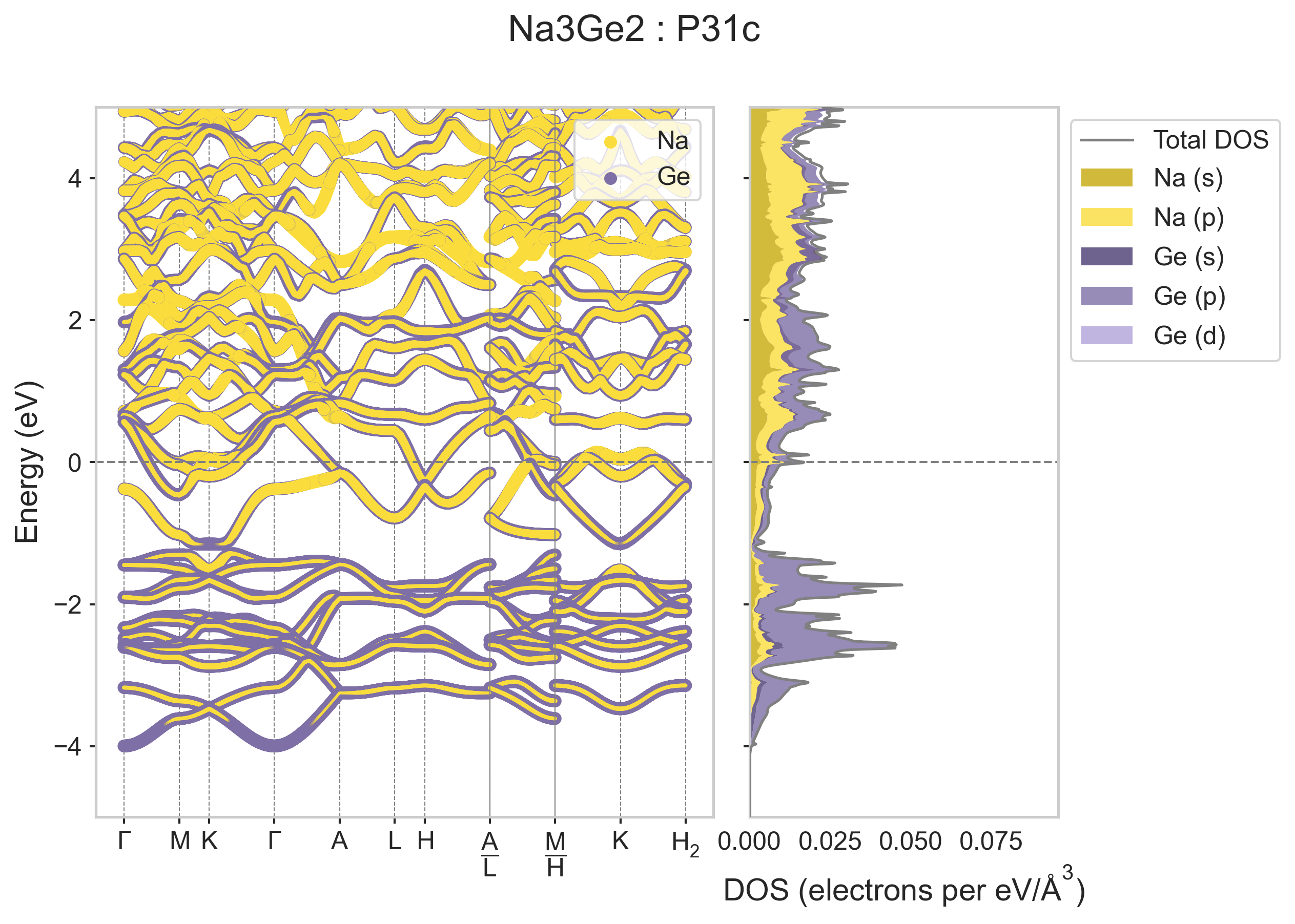}};
    \end{tikzpicture}
\caption{Electronic bandstructure of \ch{NaGe} (left) and \ch{Na3Ge2} (right) at 0 GPa projected onto the Na and Ge states and the corresponding density of states projected onto the Na s and p and Ge s, p and d states. The direct and indirect band gaps seen in NaGe are marked in red and blue respectively.}
\label{sfig:NaGe_Na3Ge2_bandstructure}
\end{figure}

\clearpage
\subsection{Motif analysis } \label{sec:motif}

\begin{figure*}[h!]
  \includegraphics[width=0.8\textwidth]{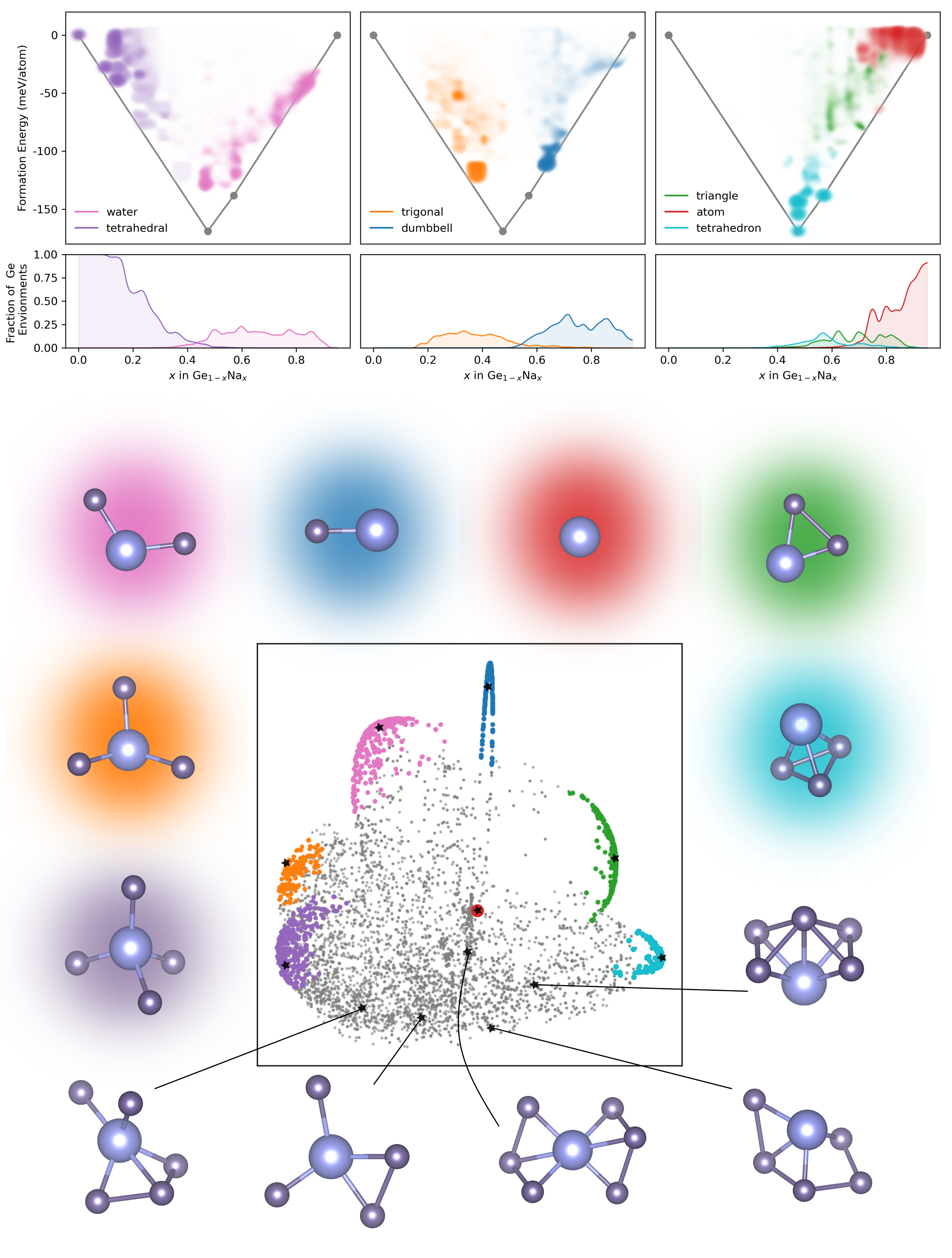}
  \caption{The Ge environments in the 0 GPa search results (structures with neagtive formation energies only) were analysed using \ac{SOAP} descriptors combined with \ac{K-PCA} for dimensional reduction (see bottom panel) and \ac{DBSCAN} for automatic clustering. The heat map in the top panels shows the fraction of Ge environments assigned to each cluster. }
  \label{sfig:motif_combined}
\end{figure*}

\clearpage
\subsection{Simulated pair distribution functions.}

\begin{figure}[h!]
  \includegraphics[width=0.8\textwidth]{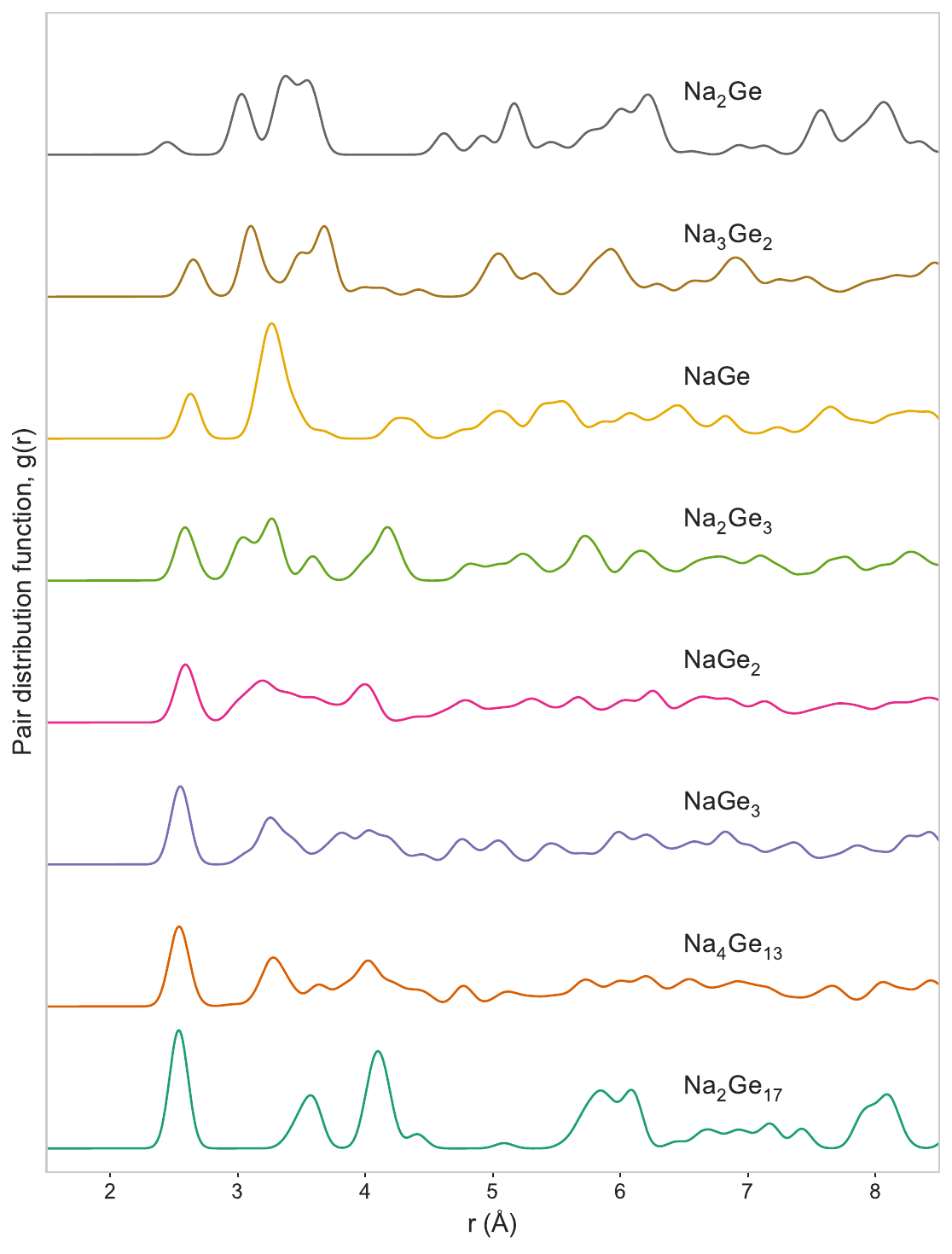}
  \caption{Computed pair distribution functions (PDF) for low energy phases at 0~GPa. All PDFs have been artificially broadened with Gaussians of width 0.1 \AA.}
  \label{fig:pdf}
\end{figure}

\begin{figure}[h!]
  \includegraphics[width=0.8\textwidth]{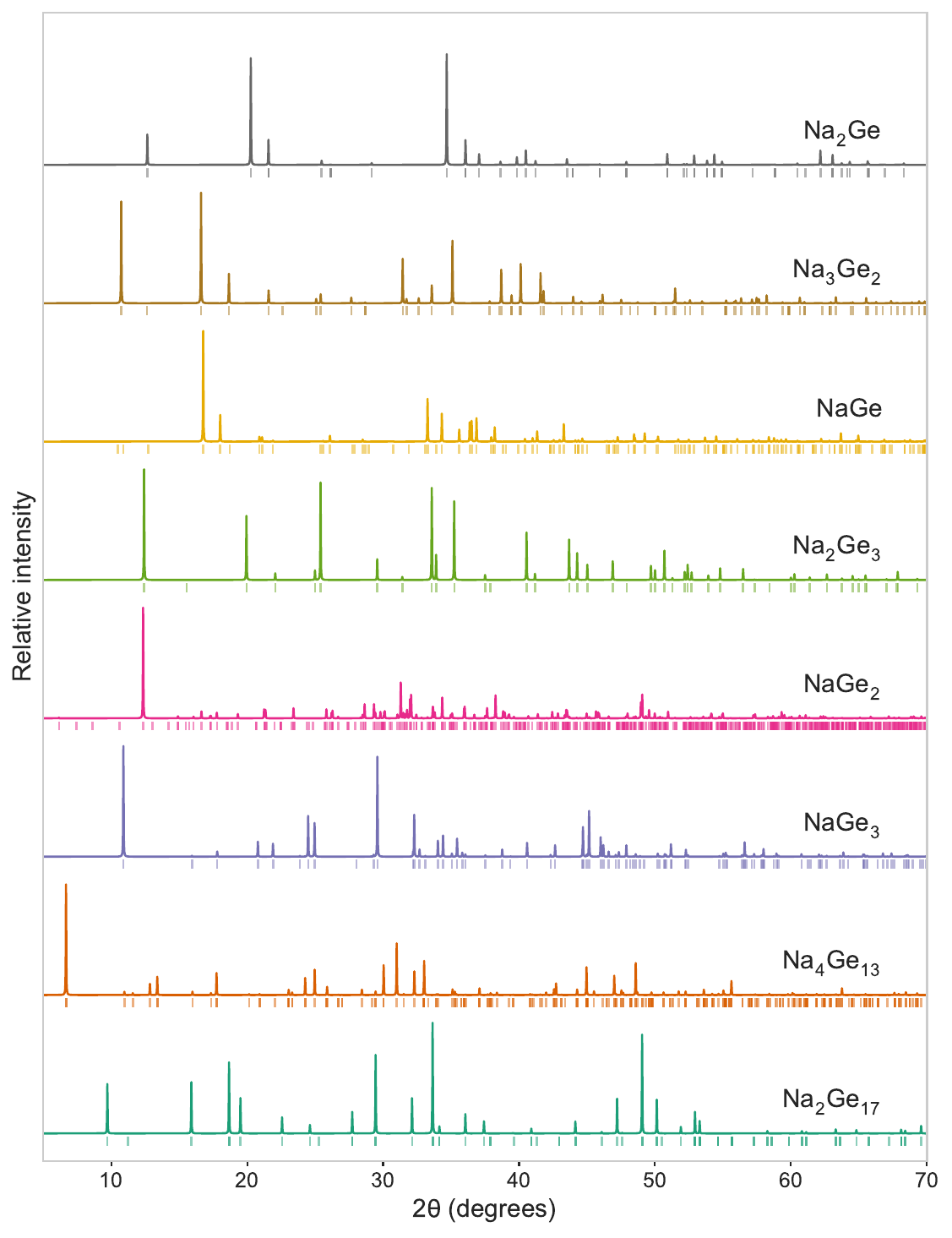}
  \caption{Simulated powder x-ray diffraction patterns for low energy phases at 0~GPa. using Cu-k$\alpha$ source.}
  \label{fig:pxrd}
\end{figure}

\clearpage
\subsection{Ge K-edge XAS Spectra} \label{sec:coreloss}
As shown in the previous Section \ref{sec:motif}, the total search space of structures on the Na-Ge convex hull can be simply divided into several smaller building blocks as shown in Fig. \ref{sfig:motif_combined}. This process is akin to the same experimental procedure of deconvolution which is commonly used to separate spectral features into atomic-level descriptors in X-Ray spectroscopies such as X-Ray Absorption Spectroscopy (XAS). By combining the localised SOAP descriptor environments with an experimental characterisation technique in a method previously described by Aarva \textit{et al.} as ``Fingerprint Spectra'' \cite{aarva2021x}, we attempt to contextualize the results of our motif separation using Ge K-edge XAS.

XAS is an expensive experimental technique, which requires obtaining time at a beamline in order to gather spectra of a given sample. Therefore, being able to predict spectra using first principles modelling prior to acquiring beam-time is a useful tool for reducing the cost of experimental structure characterisation. In addition, the `near edge' region of XAS can be probed to specifically understand the oxidation state and coordination number of a specific element using X-Ray Absorption Near Edge Spectroscopy (or XANES). XANES is both element specific and locally sensitive, and it probes both the local electronic and structural characteristics of a nucleus. By exciting core electrons into the region above the Fermi level, XANES (also referred to in the literature as NEXAFS) probes the empty states in the conduction band, thereby determining the oxidation state and local overlap of orbitals in an environment.

Ge K-edge XAS has been previously used to successfully distinguish between the rutile and quartz polymorphs of GeO$_2$ \cite{bertini2003germanium}. In addition the glass structure of SiO$_2$-GeO$_2$ glasses was determined in-part using Ge K-edge XANES \cite{majerus2008structure}. By calculating the nearest-neighbor coordination environments and angle differences between Ge-O-Ge and Ge-O-Si components, they were able to determine that the glassy mixture was comprised of a random network of GeO$_4$ and SiO$_4$ tetrahedra, for example. By calculating the Ge K-edge spectra of compounds within each of the identified `motifs', we aim to describe relevant characteristics of each motif's calculated XAS spectrum. 

To calculate the Ge K-edge XAS, we used CASTEP \cite{clark2005first} with a higher set of convergence criteria in order to ensure well-converged spectra. We used the `HARD' set of pseudopotentials, which have a smaller core radius cut-off therefore allowing more accurate results close to the nucleus where XAS is most sensitive. We used a kinetic energy cut-off of 1000\,eV and k-point spacing of 0.03 $\times$ 2$\pi$ \AA$^{-1}$. In order to reference the results to the experimental transition energy of Ge K-edge XAS, we calculated the Ge K-edge XAS for both rutile and quartz GeO$_2$ and applied a chemical shift to all of the results for our Na-Ge spectra following the method of Mizoguchi \textit{et al.} \cite{mizoguchi2009first}. The r-GeO$_2$ k-edge is at 11109.6 eV and the g-GeO$_2$ k-edge is at 11108.4 eV \cite{bertini2003germanium}.

\begin{figure}[htb!]
    \centering
    \includegraphics[width=0.9\textwidth]{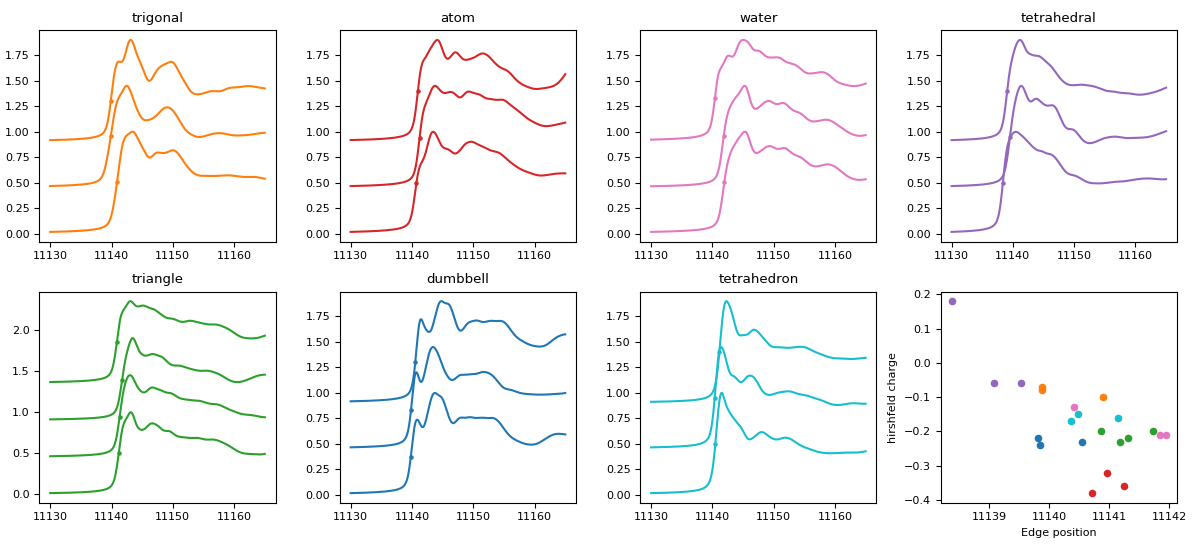}
    \caption{XAS spectra of selected structures from each motif defined in Sec. \ref{sec:motif}, calculated for the Ge K-edge of a Ge atom within the defined structural motif. The colors of each spectrum correspond to the colors in Fig. \ref{sfig:motif_combined}. The bottom right panel shows a comparison of the location of the leading edge in the Ge K-eddge XAS for each structure compared to the Hirschfield charge on the Ge atom within the cluster.}
    \label{fig:core-loss}
\end{figure}

To compare structures from each motif, we selected three representative structures from each of the motifs shown in Fig. \ref{sfig:motif_combined}. Then, in each case we calculated the Ge K-edge XAS, specifically for a Ge atom which is within the structural motif. The results for each structure are shown in Fig. \ref{fig:core-loss}, separated by the respective local environment and colored according to the same legend as in Fig. \ref{sfig:motif_combined}. 

In each case, the spectral features such as the location of the near edge, number, and intensity of peaks between different structures of the same motif are similar. The final plot in the bottom right of Fig. \ref{fig:core-loss}, shows the location of the near edge calculated as the midpoint between the onset and first peak of the spectrum for each structure, relative to the Hirschfield charge of the Ge atom. There is a general trend towards a higher energy leading edge and lower Hirschfield charge, which can be explained by the fact that it will be more difficult to remove an electron from a core state of a negatively charged atom, resulting in a higher onset energy required in the XAS.

A visual analysis of each of these spectra shows that there are several spectral fingerprints which have features that are distinct from the other structural motifs such as the double-peak feature of the dumbbell structures. The Ge K-edge shown for the three dumbbell structures in Fig. \ref{fig:core-loss} has a small pre-edge at 11141 eV and a larger broader peak at 11145 eV which is higher in energy. This is distinct from the spectra for the `trigonal' motif which have no leading edge but rather two main peaks at 11142 eV and 11150 eV. Finally, the spectra for the `tetrahedral' and `tetrahedron' results show similar leading edges (11140 eV) and peak shapes, with one main peak at 11141 eV and then a smaller peak at 11145 eV, which is expected given the similar coordination number in these two motifs. Although they are structurally distinct as shown in Fig. \ref{sfig:motif_combined}, both are four-fold coordinated Ge atoms, which likely have similar local bond lengths with Na, and therefore similar oxidation states and XAS spectra.

\begin{figure}[htb!]
    \centering
    \includegraphics[width=0.9\textwidth]{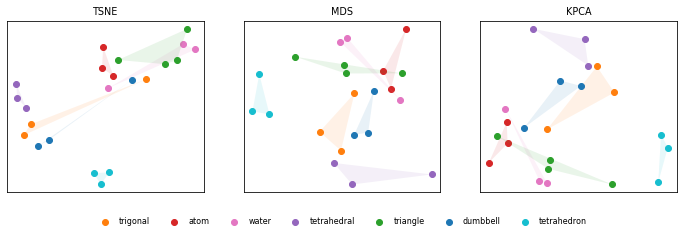}
    \caption{Principal component analysis using the Wasserstein distance between each spectrum shown in Fig. \ref{fig:core-loss}. The TSNE, MDS, and KPCA methods were all tested using an all-by-all comparison of the Wasserstein distance between each spectrum. Each point on the maps represent a single spectrum from Fig. \ref{fig:core-loss}, and are colored by the motif as defined in Sec. \ref{sec:motif}.}
    \label{fig:core-loss-pca}
\end{figure}

In order to quantitatively assess the similarity of each spectrum in Fig. \ref{fig:core-loss}, we chose to calculate the Wasserstein distance between each spectrum, and use a principal-component analysis to determine the similarity of the spectral features. The Wasserstein distances, as well as the TSNE, MDS, and KPCA principal component maps shown in Fig. \ref{fig:core-loss-pca} are all calculated using \texttt{scipy}. In all three cases, the tetrahedron motif spectra cluster separately from all of the other spectra shown, highlighting that this motif is distinct from the others shown in Fig. \ref{sfig:motif_combined}. These results provide a direct link to experimental spectroscopy, and provide guidelines for future experimentalists when assigning motifs in experimental Na K-edge XANES.

\end{document}